\documentclass[12pt]{article}

\usepackage{amssymb,amsmath,amsfonts,eurosym,geometry,ulem,graphicx,caption,color,setspace,sectsty,comment,footmisc,natbib,pdflscape,subfigure,array,theorem, mathpazo, xcolor, tabularx, cellspace, setspace, booktabs, makecell, fancyhdr, enumitem, adjustbox, rotating, tikz, pdfpages, mathtools, cancel}

\usetikzlibrary{positioning,calc,arrows,arrows.meta}
\definecolor{fv}{HTML}{eff1f5}
\definecolor{mycyan}{HTML}{00bbdd}
\usetikzlibrary{decorations.pathreplacing,calligraphy}
\usetikzlibrary{trees}

\usepackage[toc,page,header]{appendix}

\RequirePackage{titletoc}

\usepackage{threeparttable}
\usepackage{multicol}

\usepackage{float}
\newtheorem{prop}{Proposition}

\newtheorem{proof}{Proof}
\newcommand{\sym}[1]{\textsuperscript{#1}}

\theorembodyfont{\rm}

\newcolumntype{L}[1]{>{\raggedright\let\newline\\arraybackslash\hspace{0pt}}m{#1}}
\newcolumntype{C}[1]{>{\centering\let\newline\\arraybackslash\hspace{0pt}}m{#1}}
\newcolumntype{R}[1]{>{\raggedleft\let\newline\\arraybackslash\hspace{0pt}}m{#1}}

\usepackage [english]{babel}
\usepackage[utf8]{inputenc}
\usepackage [autostyle, english = american]{csquotes}
\MakeOuterQuote{"}
\usepackage{natbib}
\usepackage{bbm}
\usepackage{comment}
\usepackage{nameref}

\usepackage[hyphens]{url}
\usepackage[hidelinks]{hyperref}
\definecolor{softblue}{HTML}{1C46C0}
\definecolor{softred}{HTML}{B54747}
\definecolor{psered}{HTML}{DC0C47}

\hypersetup{%
 colorlinks = true,
 linkcolor = psered,
 citecolor = psered,
 urlcolor = psered
}

\begin{document}

\begin{titlepage}
\title{\fontsize{19pt}{19pt}\selectfont \sc{\bfseries Critical Thinking Via Storytelling: Theory and Social Media Experiment}}
\vspace{2cm}

\author{Brian Jabarian\thanks{Paris School of Economics. Contact: \href{brian@jabarian.org}{brian@jabarian.org}.} \hspace{35pt} Elia Sartori \thanks{Center for Studies in Economics and Finance and Università degli Studi di Napoli Federico II.}}

\date{\today\footnote{We are indebted to Roland Bénabou for his guidance. We are grateful for the comments of Tore Ellingsen, Nicolas Jacquemet, Yves Le Yaouanq, Dan McGee, Pietro Ortoleva, Jean-Marc Tallon, Marie-Claire Villeval, Leeat Yariv, and Sam Zbarsky. We are grateful for the research assistance provided by Christian Kontz, Andras Molnar, Alessandro Sciacchetano, and Alfio De Angelis. We are grateful for the participation of psychologists from the Department of Psychology at Princeton University. We thank seminar audiences at Bologna, PSE. We obtained the Princeton IRB Approval \#12995 on June 12, 2020.
This paper was partly written while Brian visited the Department of Economics at Princeton University and the Kahneman-Treisman Center for Behavioral Science and Public Policy in 2018-2020. He thanks their hospitality. He also acknowledges financial support from the Paris School of Economics, Sorbonne Economics Center, and the Forethought Foundation for his visit to Princeton, Grant ANR-17-CE26-0003, and Grant ANR-17-EURE-001. \textit{Full Acknowledgment to be added}.}}
\end{titlepage}
\maketitle


\begin{abstract}
\noindent 
In a stylized voting model, we establish that increasing the share of \textit{critical thinkers} -- individuals who are aware of the ambivalent nature of a certain issue -- in the population increases the efficiency of surveys (elections) but might increase surveys' bias. In an incentivized online social media experiment on a representative US  population  ($N = 706$), we show that different digital storytelling formats -- different designs to present the same set of facts -- affect the intensity at which individuals become critical thinkers. Intermediate-length designs (Facebook posts) are most effective at triggering individuals into critical thinking. Individuals with a high need for cognition mostly drive the differential effects of the treatments. 

\bigskip
\end{abstract}

\newpage
\onehalfspacing
\section{Introduction}\label{intro}

Individuals often hold conflicting attitudes towards complex issues, known as attitudinal ambivalence (\cite{kaplan1972}). Such attitudes are particularly prevalent when an issue presents both positive and negative aspects and when the available hard evidence is inconclusive. For example, the debate on whether society should implement strict digital policies to protect users' privacy at the expense of less innovation may be perceived as a dilemma by individuals since the arguments available do not support unambiguously one side or the other. This dilemma of balancing privacy and innovation can lead to ambivalent attitudes as different facts and arguments can support both sides.
Such dilemma situations, with digital content produced through generative Large Language Models and Artificial Intelligence hardly easily verifiable with clear-cut evidence, are likely to become prevalent on digital platforms. 

When faced with complexity and ambivalence, individuals may respond through critical thinking (\cite{kahneman2011thinking}).\footnote{Individuals may also respond based on emotional bias and motivated reasoning (\cite{kunda1990case}, \cite{benabou2015economics}); however, in this paper, we focus on critical thinking.}  Critical thinking goes beyond simple cost-benefit analysis and involves a metacognitive process of becoming aware of one's ambivalent attitude toward a dilemma and evaluating the various perspectives that underlie each attitude (\cite{halpern2013thought}). Through this measurable process (\cite{list2022enhancing}), individuals can overcome their ambivalence and pass from holding \textit{raw preference} to forming \textit{stable preference}. For decision-makers relying on surveys and elections, stable preferences represent more reliable data than raw preferences. 

Consider public figures or organizations whose social image or economic returns depend on the approval of their public stance on a particular issue. These principals must predict the public stance their audience expects them to take because a public endorsement is a reputational commitment and a "focusing event" that nudges agents to think critically about their raw preferences and establish a stable preference, which might go against the principal’s stance. Hence, the principal should estimate  agents’ stable preferences before taking a stance, minimizing the risk of lasting backlash. Suppose such an estimate is based on a poll. Then, its quality hinges on agents reporting their stable preference, for which it is necessary that they have formed them in the first place.\footnote{One might conceive a strategic voting setting  where agents misreport their stable preference even after they have formed one. We think this concern is second-order for the application we have in mind, so we assume that forming and reporting a stable preference to go hand in hand. } 

Eliciting stable preferences might also be important  for an institutional principal, i.e., a policy-maker responsible for designing an economic policy on a societal issue that presents a binary dilemma (for instance, egalitarian versus freedom). 
The principal can choose from a larger (continuum) set of policies. The optimal policy is a function of the distribution of stables preferences, i.e., the share of individuals that would prefer one alternative over another at the end of the critical thinking process. For concreteness, think of a situation where the optimal amount of a social welfare program depends on the share of citizens who support an egalitarian society against a free-market one. 
The policy-maker must adopt a policy aligned with one of two opposing worldviews. At most, they would know their stable preference on the issue but risk imposing it on the remaining population. In this interpretation, the policy-maker uses the citizens' stable preferences distribution as a normative  criterion.   
Hence, here also, it is in the institutional principal’s interest to anticipate (and incentivize) the formation of agents’ stable preferences before making her decision.   

In summary, we posit a principal who cares about the distribution of stable preferences either because she fears her action will cause a backlash if they are too distant from the target or because she is using such distribution "for lack of anything better" as a proper normative criterion for social aggregation of preferences. Elections would be efficient if all individuals reported their stable preference at the poll, allowing a perfect tracking of the relevant unknown. However, individuals find their stable preference only at the end of a  critical thinking process which not every agent is able or willing to undertake. Also, eliciting such a process requires novel experimental methods since standard survey methods fail to classify the types of agents. 

In such an ambivalent environment, hard information is insufficient to resolve agents' ambivalent attitudes, hence, underlying the role of ``storytelling formats'' and the cognitive styles of agents in forming stable preferences. In an ambivalent environment, where there is a lack of consensus on important issues, the role of media is not only to provide objective information but also to push individuals to realize the ambivalent nature of the issue. These "focus events" are not based on what is traditionally known as "information" but rather on stylized and partial facts communicated through various media formats such as Facebook posts, news articles, and tweets. These facts, while potentially irrelevant to the judgment of value, can still impact the speed at which individuals recognize the ambivalence of the issue.

This paper refers to such a " storytelling format" as the packaging (UX design) of information, which is crucial in triggering individuals toward critical thinking. Related to \cite{aragones2005fact}, they perform 'fact-free \textit{self}-learning'. A well-reasoned essay may trigger individuals to become aware of the issue's complexity, while a sequence of bombarding facts may induce the same outcome. The quantity and quality of information presented and the individual's cognitive style play a crucial role in this process. In our stylized digital economy, the media is a nudging device, triggering individuals to transition into critical thinking. This is because once an individual realizes they are ambivalent about an issue, no "objective fact" can drive their preference. Instead, critical thinking, in which different worldviews are weighted, is necessary to form a stable preference. Therefore, we will adopt the terminology "stories" and "storytelling formats" to refer to "media content" and "media format," respectively. How a problem is presented, from a shallow tweet-storm to a well-reasoned newspaper article, can affect how quickly individuals realize ambivalence about an issue.

\paragraph{Aims of the paper.} In this paper, we study the role of storytelling formats in shaping this process, with the idea that realizing the ambivalent nature of an issue (i.e., starting the process) is an event whose likelihood depends on the storytelling format associated with the informational environment. Hence, the question we address is the following: do storytelling formats affect the efficacy of polls at eliciting stable preferences by impacting an individual's critical thinking process? Besides, can  agents' cognitive styles explain this effect? To address this question, we need a theoretical model to formalize the awareness of ambivalence within a population affects the predictive power of elections and an empirical demonstration that different storytelling formats push individuals into critical thinking at different rates. This paper accomplishes both tasks and provides an affirmative answer: storytelling formats matter for the efficiency of elections in theory and practice. 

\paragraph{Theoretical setting.}

We construct a simple but novel theoretical model in which the intensity, $\lambda$, at which ace
individuals in a (large) population enter critical thinking is a relevant welfare measure. The principal uses the outcome of elections to guess the distribution of stable preferences inside the population. The raw preference (or stereotype preference), $x_{S}$, is drawn from a perturbation of the distribution of stable preferences -- the object the principal wants to elicit. Although the reported preference, $x_{A}$, is stochastic (given the stable preference) even during the critical thinking phase, allowing for ``reasoned indecisiveness,''\footnote{The baseline model, which can be thought of as a simplified version of a fully identified three-state model, is presented in the Appendix and delivers no significant qualitative difference in the results, wherein individuals complete their critical thinking process and finally discover (and report) their stable preference.} removing the stereotype component orthogonal to the stable preference improves the ``quality'' of the reported preference. For this reason, the average reported preference becomes more informative about the average stable preference the more agents successfully perform critical thinking. Since such informativeness coincides with the welfare of a principal that ``plays'' the estimator obtained in the poll, we posit that $\lambda$ is indeed a welfare measure.\footnote{Comparative statics are ambiguous if we consider an institutional principal that has to act according to the election outcome and, therefore, must consider election bias.}  

\paragraph{Social Media Experiment.}
Based on this sharp theoretical result, we conduct an online experiment to demonstrate that storytelling formats affect $\lambda$ and, therefore, the efficiency of elections. To do so, we present a representative sample with a ``relatively new'' ambivalent issue, namely, digital privacy. Accordingly, the debate is whether we should remunerate users for data sharing. We randomly assign subjects to one of four storytelling format treatments: \textsc{twitter}, \textsc{facebook}, \textsc{newspaper}, or \textsc{partisan twitter}. We maintain the same content (i.e., we give agents the same selection of information) but alter its storytelling format. We elicit pre- and post-treatment awareness of attitudinal ambivalence in two different -- although both incentivized -- ways and test whether the storytelling format alters the likelihood of the agent moving into critical thinking. We find that intermediate-length stories (the \textsc{facebook} treatment) are significantly more effective at nudging individuals into critical thinking. We then investigate whether individuals' cognitive styles can (at least partially) explain this heterogeneous response to the storytelling format. To this end, we split our sample along two cognitive traits that we also elicit throughout the experiment using standard measures: need for cognition (\cite{cacioppo1982need}) and cognitive flexibility (\cite{martin1995new}). We find that most of the heterogeneity in the response is driven by individuals with a high need for cognition. At the same time, the cognitive flexibility score is irrelevant in determining responsiveness to critical thinking.

\paragraph{Literature Review.}

\paragraph{\it Critical thinking and non-dogmatism.}
In its normative interpretation, our behavioral model relates to \cite{millner2020}. In his model, non-dogmatic social planners are insecure in their future time preferences and entertain the possibility of endorsing different ones. Millner's work is reminiscent of our normative interpretation in which the planner takes the distribution of stable preferences as the ``truth'' when it comes to a dilemma and is only concerned that an election would give a partisan view because not everyone experiences moral uncertainty.

\paragraph{\it Simplistic Rhetoric.}
Research has shown that political discourse has become increasingly simplified over time (\cite{jordan2019examining}) and that this trend intensifies during election periods (\cite{tetlock1981pre},   \cite{thoemmes2007integrative}, \cite{conway2012does}). Our model suggests that this simplification can lead to awareness for individuals. Additionally, our model also suggests that simplification can lead to a skewing of preferences among individuals in a stereotypical phase of the process, where one argument may be seen as more persuasive due to overconfidence \cite{ortoleva2015overconfidence} and the belief that one's experiences are more informative about policy than they are. This highlights the importance of considering the effects of simplistic rhetoric in political discourse and the importance of individuals being aware of their ambivalence towards an issue.

\paragraph{\it Predictive Power of Elections.}
As demonstrated in the seminal work of \cite{feddersen1997voting}, there are instances where many voters effectively aggregate information, resulting in an equilibrium outcome that is fully information-equivalent. However, preference heterogeneity can impede a voting procedure from effectively aggregating individual voters' information (\cite{kim2007swing}; \cite{gul2009partisan}; Bhattacharya, 2013; Acharya, 2016; Ali et al., 2018). This literature highlights that when voters have divergent preferences and incomplete information about the state of nature, they may collectively choose an outcome that is less favorable for society or preferred by a minority. The central concern in these papers is strategic voting, an issue not present in our analysis. In our model, every citizen votes for their current preference. Still, the extent to which this preference accurately reflects the payoff-relevant stable preference depends on the citizen's cognitive state, which is influenced by politics. Political  can be seen as a method of making citizens view their stable preference as private information that an election aims to uncover.

\paragraph{\it Social Media and Welfare.}
Our paper contributes to expanding an already vast literature that has been focusing more and more on how a specific \textit{format} over the last decade (i.e., social media could shape agents' political behaviors such as voting). \cite{gorodnichenko2021} empirically showed that using bots in social media has a role in influencing public opinion. \cite{munir2018social} provided evidence of the impact of social media in shaping the voting behavior of Scottish youth during the Scottish Independence Referendum of 2014. \cite{falck2014} studied the impact of more general information disseminated over the Internet on voting behavior and found that it influenced it.


\paragraph{Structure of the paper.}
Section \ref{theory} describes the behavioral model and its main positive and normative results. Section \ref{experiment}
 elaborates on the experimental design. Section \ref{mainresults} elaborates on the empirical results. Section \ref{conclusion} concludes and discusses possible caveats and extensions of the model and experiment. 

\section{Theoretical Setting}\label{theory}

We consider a stylized political choice setting. The utility is the distance between the political action and a target determined by the distribution of the preferences individuals hold if they had completed a \textit{critical thinking} process. This parameter is unknown at the onset and can only be estimated using the outcome of a poll held at some time $t$, when (a part of) citizens might still not have completed their process. 

The main aim of this section is to establish -- Proposition \ref{prop:Main-Result} -- that the intensity at which citizens complete their critical process is a relevant welfare measure: whatever the time of elections is, higher intensity increases the information content of elections. This gives our exercise of estimating the intensity associated with different propaganda formats -- the objective in the experiment presented in Section 3 -- to have normative content. However, we also show that the main message relies on assuming that we have a principal who can freely manipulate the  outcomes of the poll when taking her action (which we refer to as $P$ositive principal). If the election outcome constrained her action, as is most likely the case for an $I$nstitutional principal, then a bias-precision trade-off emerges, which makes the comparative statics ambiguous. 

We present the two welfare benchmarks (corresponding to the two types of principals discussed in the introduction) and voters' critical thinking process separately. We then put them together to derive our main result and some important caveats that shed light on the assumptions needed to get the unambiguous comparative statics.

\subsection{Principal: Two Welfare Benchmarks}

The relevant unknown is the distribution of stable preferences in a large (continuum) population,
namely the share $p\in\left[0,1\right]$ of individuals who would prefer outcome $1$ to outcome $0$ after they complete their critical thinking process. Welfare realizes the distance
between the social action $a$ and its target $p$\footnote{Although the space of preferences --- individuals' resolution of the moral dilemma --- is binary, the policy space is continuous. This corresponds to a situation where the planner can fine-tune the policy to the \textit{distribution} of individuals' preferences. the example in the introduction of choosing the size of the welfare program based on the share of people that hold an egalitarian (rather than a free-market) view fits this story. A different specification would  $a^\star=\mathbb{I}[p>\frac{1}{2}]$ (binary action space) delivers similar insight but is less tractable. }:

\[
W\left(a,p\right)=-\left(a-p\right)^{2}
\]
Ex-ante, $p$ is unknown and drawn from a normal distribution $p\sim\mathcal{N}(\mu,\sigma)$; absent the information from the election, the principal would then choose $a=\mu$ and get value $-\sigma{^2}$.\footnote{The normality assumption gives tractable conditional expectations and closed-form welfare. It is inconsistent with the compact support $[0,1]$. The analysis with ex-ante uniform $p$ (and $p_S$) is algebraically more involved but does not change the qualitative results. For tractability, we keep the normal setup, implicitly assuming that $\sigma$ is "small enough" that the mass outside $[0,1]$ is negligible.} Before choosing $a\in\left[0,1\right]$, the principal observes the proportion $\bar{p}$ of agents that report to prefer alternative $1$. We call
$\bar{p}$ the \textit{election outcome}. and consider two types of principals that differ in the use they can make of this information.

\paragraph{Positive Principal.}
A $P$ositive principal, for which the election outcome is \textit{not} binding, namely who can choose any $a\in[0,1]$ no matter the realization of $\bar{p}$.
The Positive principal uses the election outcome and his knowledge of the critical thinking process inside the population (next Section) to
estimate $p$. His optimal action is the conditional expectation 
\begin{align*}
    a^{\star}=\hat{p}\coloneqq\mathbb{E}[p|\bar{p}]
\end{align*}

that achieves value 
\begin{equation}
W_{P}=-\mathbb{E}\left[\left(\hat{p}-p\right)^{2}\right],
\end{equation}\label{eq:WelfPos}
equal to the dispersion of the conditional mean $\hat{p}$ around $p$.\footnote{The expectation operator $\mathbb{E}$ integrates under the joint distribution of $p,\bar{p},\hat{p}$, which depend on the voting behavior and citizens' critical thinking process and that we derive in the next section.}
Connecting to the discussion in the introduction, one can think of such principals as public figures (e.g., multinational firms or social influencers with reputational concerns) that need to take
a stance on an ambivalent issue. They privately run a poll (say, by asking a polling agency) and use its outcome as they wish to fine-tune their statement. Payoff depends on the (distribution of) stable preferences because the statement acts as a "focusing event" that pushes the relevant population into critical thinking: the preferences individuals judge the principal on are (potentially) different from those they report at the poll.

\paragraph{Institutional Principal.}
An $I$nstitutional principal, whose action is constrained to be $a=\bar{p}$, which achieves value
\begin{equation}\label{eq:WelfNor}
W_{I}=-\mathbb{E}\left[\left(\bar{p}-p\right)^{2}\right]
\end{equation}
One can think of such principals as democratic institutions that must comply with the election outcome (say, by empowering a parliament whose composition is proportional to $\bar{p}$).\footnote{In this context, the interpretation of $p$ differs. Rather than focusing on the potential backlash from stable preferences, we envision an institutional principal considering $p$ as a normative criterion for aggregating social preferences about an ambivalent issue. Essentially, the distribution of preferences of individuals who have undergone the critical thinking process determines the "right thing to do".}  Via the standard decomposition, we get that 
\begin{equation}
W_{I}=W_{P}-B,\label{eq:WelDec}
\end{equation}
 where 
\[
B=\mathbb{E}\left[\left(\bar{p}-\hat{p}\right)^{2}\right]>0
\]
is the bias of election, representing how the average reported preference differs systematically from the stable ones. A principal $P$ who can correct for such social tendencies only suffers from
the dispersion of the estimator $\hat{p}$ around the parameter $p$, whereas the principal $I$ is also concerned by the bias of the election since they cannot correct them.

\subsection{Agent: Cognitive and Voting Processes}

Each individual is characterized by a stable preference 
\[
y\sim Ber(p)
\]
where $p$ is the welfare relevant unknown that the principal wants
to match. For example, an individual with $y=1$ has a stable preference for alternative $1$.
If asked at a poll, however, individuals do not necessarily report
their stable preference. This is because the stable preference is
"discovered" at the end of  a \textit{critical thinking process} that individuals undergo.
\paragraph{The Cognitive Process.} Agents transition through two critical thinking \emph{states}
$\left\{ S,A\right\} ,$ where $S$ means \textit{Stereotype} and
$A$ means \textit{Awareness}. We assume that the critical thinking process follows a simple dynamic in continuous time: all individuals
start at $t=0$ in state $S$ and, independently of $y$ (and other
voting parameters), they transition to the absorbing state $A$ with intensity
$\lambda\in\left(0,\infty\right)$. Therefore, at time $t$ there will
be a fraction 
\[
\eta_{S}=\exp\left\{ -\lambda t\right\} 
\]
 of agents that are still $S$tereotypes and $\eta_{A}=1-\eta_{S}$
that transitioned to Awareness.\footnote{The assumption that $A$ is an absorbing state, with no transitions from $A$ to $S$, captures the idea that awareness is an irreversible process. A straightforward extension of the model prevents a scenario where all individuals eventually reach state $A$: a constant fraction $\nu < \lambda$ exits the economy and re-enters in the awareness state $S$. Qualitative results would be unchanged as the associated share of stereotypes: 

 $$\eta_S(t) =\frac{\nu}{\lambda}+\exp\left(-\lambda t\right)\left(1-\frac{\nu}{\lambda}\right)$$

would still be decreasing in $\lambda$,$t$.} 
The parameter $\lambda$ is key for our analysis. It represents the intensity with which individuals realize the issue at hand is ambivalent. We estimate its value by estimating in our experiment for different propaganda formats: the idea is that the way news is presented has a role in determining the speed at which individuals move into $A$ -- and that such difference interacts
with other cognitive abilities. 

\paragraph{Voting Behavior.} 
We denote $x$ the preference that individuals report at the polls and assume it  depends as follows  on the stable preference $y$ and on the stage of the critical thinking process $\{S,A\}$.
Before realizing the issue is ambivalent, the reported preference $x_S$ is 

\[
x_{S}\left|y\right.=\begin{cases}
Ber\left(p_{S}\right) & \text{w.p. }\beta\\
y & \text{w.p. }1-\beta
\end{cases}
\]

In words, $x_S$ is equal to the stable preference with probability $\beta\in[0,1]$, while complementary probability is drawn from a distribution of stereotypical preferences $p_{S}\sim\mathcal{N}(\mu,\sigma)$, independent of $p$. Since we still have a parameter $\beta$ that drives the correlation between average stereotypes and stable preferences, the independence assumption is innocuous. It only requires the formation of stereotypical preferences involving factors not solely related to $p$.\footnote{The identical distribution of $p$,$p_S$ is instead for tractability alone. Most derivations in the Appendix utilize non-identically distributed normal variables ($\mu_p$, $\sigma_p$, $\mu_{p_S}$, $\sigma_{p_S}$). Specifically, the condition $\mu_p \neq \mu_{p_S}$ illustrates a scenario in which the principal is aware that stereotypical preferences exhibit systematic bias, potentially due to the ease of presenting superficially persuasive arguments in favor of one alternative. This, along with the relaxation of other symmetry assumptions inherent in our model, is further explored in Section [to be added].} For example, if $\beta=\eta_S=1$, corresponding to a poll held at $t=0$ and where stereotypes are independent of stable preferences, then the election outcome $\bar{p}=p_S$ is uninformative about $p$.

The  preference reported by individuals in $A$ loses its dependence on the nuisance parameter $p_{S}$ and becomes a function of the stable preference alone, 

\[
x_{A}\left|y\right.=\begin{cases}
y & \text{w.p. }\xi\\
1-y & \text{w.p. }1-\xi
\end{cases}
\]

The parameter $\xi\in[\frac{1}{2},1]$ allows for a situation where citizens realize the issue is ambivalent but still have not found their stable preference. We think of our two-stage critical thinking process as a reduced form of a fully identified three-stage process -- detailed in the Appendix -- where $A$ is an intermediate stage where agents have realized the ambivalent nature of the issue but have not formed their stable preference yet -- i.e., they are in a phase of normative uncertainty. In this interpretation, the case $\xi=1$ corresponds to a situation where individuals discover their stable preference immediately after realizing the ambivalence of the issue, while $\xi=\frac{1}{2}$ a situation of permanent indecisiveness of $A$ individuals. Notice that if all individuals were in $A$ state (i.e., a $t\to\infty$ poll), then the election outcome would be $\bar{p}=\xi\cdot p+(1-\xi)\cdot(1-p)$, which whenever $\xi>\frac{1}{2}$ is a strictly monotonic (hence invertible) function of $p$. Since the election outcome contains all information about $p$, then $\hat{p}=p$, and the $P$ositive principal always chooses the correct action.\footnote{Institutional principal still needs to consider the attenuation bias driven by $A$'s indecisiveness.} In general (i.e. for interior shares $\eta$), the election outcome is given by:

\begin{equation}
\overline{p}=\eta_{S}\left(\beta p_{S}+\left(1-\beta\right)p\right)+\eta_{A}\left(\xi\cdot p+(1-\xi)\cdot(1-p))\right)
\end{equation}

and the parameter $\xi$ affects the $P$ositive welfare too. We can now use the (joint) normality assumption to write $\bar{p}$ as well as the conditional expectation $\hat{p}$ as a linear function of the fundamental unknowns $p,p_S$, that is  

\[
\bar{p}=\alpha_{0}+\alpha_{1}\cdot p+\alpha_{2}\cdot p_{S}
\]
\[
\hat{p}=\gamma_{0}+\gamma_{1}\cdot p+\gamma_{2}\cdot p_{S}
\]

where the loadings $\boldsymbol{\alpha,\gamma}$ are functions of the structural parameters $\vartheta=[\beta,\xi,\boldsymbol{\mu,\sigma}]$ and the critical thinking process statistic $\eta$ as detailed in the Appendix. Moreover, once we specify the joint normal expectation operator, we can compute (the evolution of) both positive and institutional welfare \ref{eq:PWelf}-\ref{eq:NWelf} in closed form and arrive at our main result. 

\begin{prop}
\label{prop:Main-Result}
$i)$ For all values of the structural parameters $\vartheta$, $W_{P}$ is increasing in $t$ and $\lambda$. 

$ii)$ $W^{I}$ has non-trivial comparative statics in $\lambda,t$. If $\beta<1-\xi$, then it is monotonically increasing;  
if $\beta>\frac{\left(1-\xi\right)\left(\left(1-2\mu\right)^{2}+4\sigma^{2}\right)}{2\sigma^{2}}$
then it is monotonically decreasing; else it grows local to $t=0$ (resp. $\lambda=0$) up to a finite time $t^{\star}$ (finite intensity $\lambda^\star$) then eventually decreases.
\end{prop}

Figure [to be added] gives a graphical representation of the results collected in Proposition \ref{prop:Main-Result}, which we now discuss. Point $i$ establishes that if the  principal  knows the value of the structural parameters $\vartheta$ and can utilize the outcome of elections without constraints, then the faster individuals move into critical thinking (the higher $\lambda$), the higher the efficiency of elections. Indeed in all the plots of the figure, we observe that positive welfare $W^P$ is increasing in time.\footnote{As the proof relies on $W^P$ being decreasing in the share of stereotypes eta s, the same graph would be obtained if we fix the time and let $\lambda$ vary. The bottom-right panel explains the dynamics for high and low $\lambda$.} The reason behind this result is simple to grasp: as fewer and fewer individuals are $S$, the election outcome $\bar{p}$ becomes less and less dependent on the nuisance unknown $p_S$ which confounds the inference of the welfare relevant unknown $p$.\footnote{Indeed, this result does not require the normality assumption but can be directly deduced by the expression of $\bar{p}$.} This is an important result for our analysis as it establishes that $\lambda$ is a welfare measure in a well-definite sense in our setting. 

However, in point $ii)$, we also hint at a potential limitation of such a result in case the principal is constrained to act according to the election outcome due to the (potentially perverse) effect that moving into critical thinking has on the bias of the election. The most paradoxical result -- the condition that if $\beta$ is large enough, institutional welfare is actually \textit{decreasing} in $\lambda$ -- has a natural explanation. When $\beta$ is large, then stereotypes are strong predictors of the stable preference (in the extreme where $\beta=1$, all stereotypes vote $y$ despite failing to realize the ambivalent nature of the issue),\footnote{In our setting there is no intrinsic social value for being critical thinkers so if all agents get their stable preference right we have perfect elections. However, a related phenomenon studied by \cite{bernheim2021theory}, ``mindset flexibility'' might have social benefits beyond increasing the accuracy of elections. The challenge for us is to derive $\lambda$ as a welfare measure even without a direct beneficial effect from critical thinking.} hence moving in the $A$wareness state indecisiveness and associated attenuation bias -- case $\xi<1$ -- pushes $\bar{p}$ away from $p$ and thus reduces efficiency. This seems -- at least to us -- a pathological case since it requires. but is useful to highlight the potential role of the bias. For this reason, we further investigate conditions under which the two rules coincide, that is, whether there is a level of $\eta$ such that "by divine coincidence" the loadings $\boldsymbol{\alpha}=\boldsymbol{\gamma}$ so the positive and institutional principal have the same action rule -- and hence the same value at potential limitations of this interpretation; we indeed show that if $\xi=1$ or $\beta<\frac{1}{2}$, the condition for $W^I$ monotonically increasing is vacuously satisfied.

Solving the system of equations $\alpha\equiv\gamma$ gives a share of stereotypes $\eta^{\star}$ such that the two coincide. Hence, there exists an interior time where the average reported preference is unbiased
for $p$. Formalizing this result we obtain:
\begin{prop}
If there is no bias in the stereotype pool and $\beta$ is large enough, i.e. if \[\mu_{p}=\mu_{p_S} and \frac{1-\beta}{\beta}<\frac{\sigma_{p_S}^2}{\sigma_p^2}\], then there exists a finite time $t^\star$ such that  $B\left(t^{\star}\right)=0$.
If, in addition, \[\xi=1 then t^{\star}=-\frac{1}{\lambda}\log\left(\frac{\sigma_{x}^{2}}{\beta\left(\sigma_{x}^{2}+\sigma_{y}^{2}\right)}\right)\] with immediate comparative statics.
\end{prop}

\begin{figure}[H]
    \centering
    \includegraphics[scale=0.15]{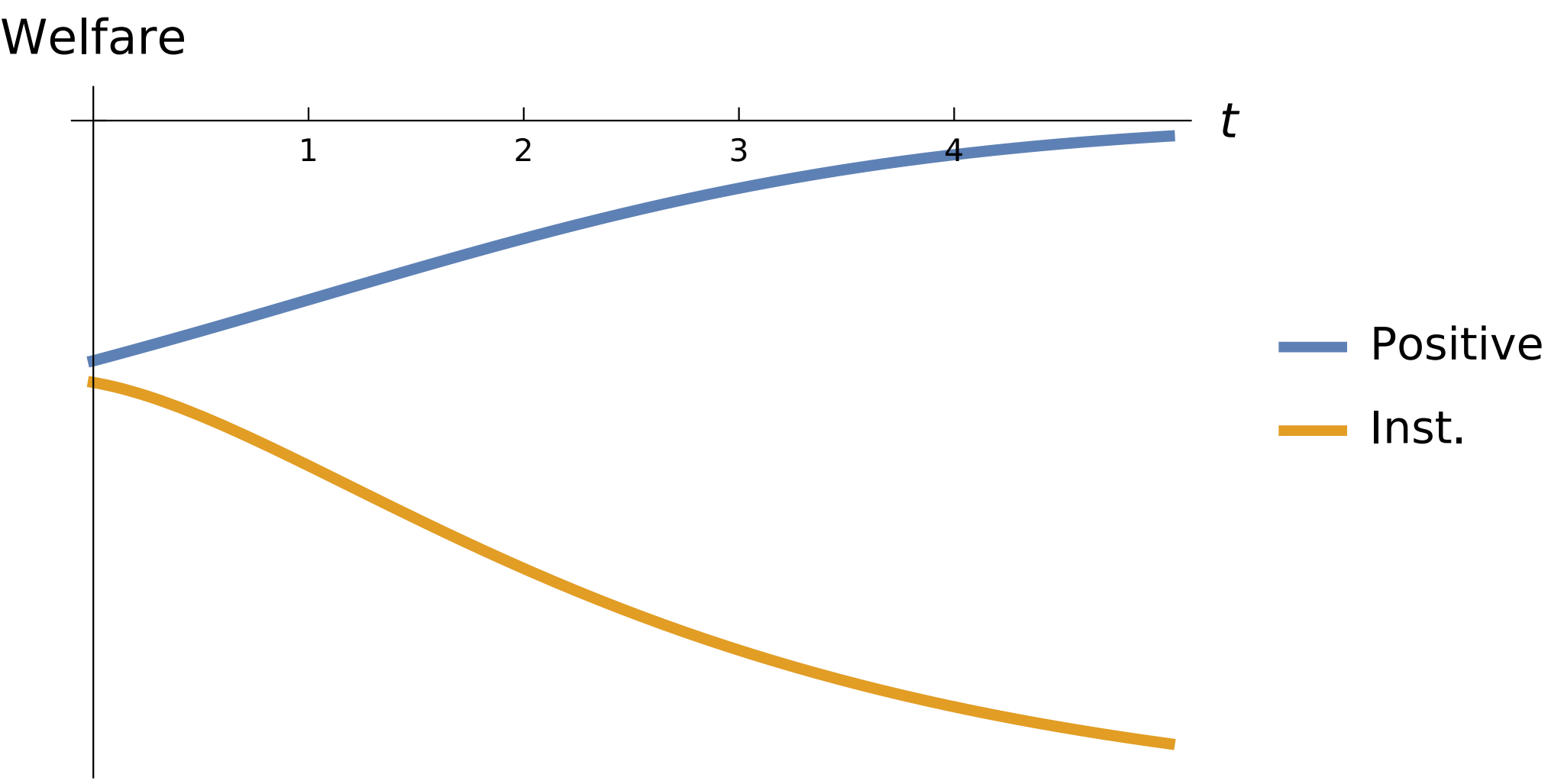}
   \caption{$\beta<1-\xi_{NU}$ $\Rightarrow$ $\eta^{\star}=0$ $\Rightarrow$ \text{Inst. Welfare is decreasing}}
    \label{fig:plot3}
\end{figure}

\begin{figure}[H]
    \centering
    \includegraphics[scale=0.15]{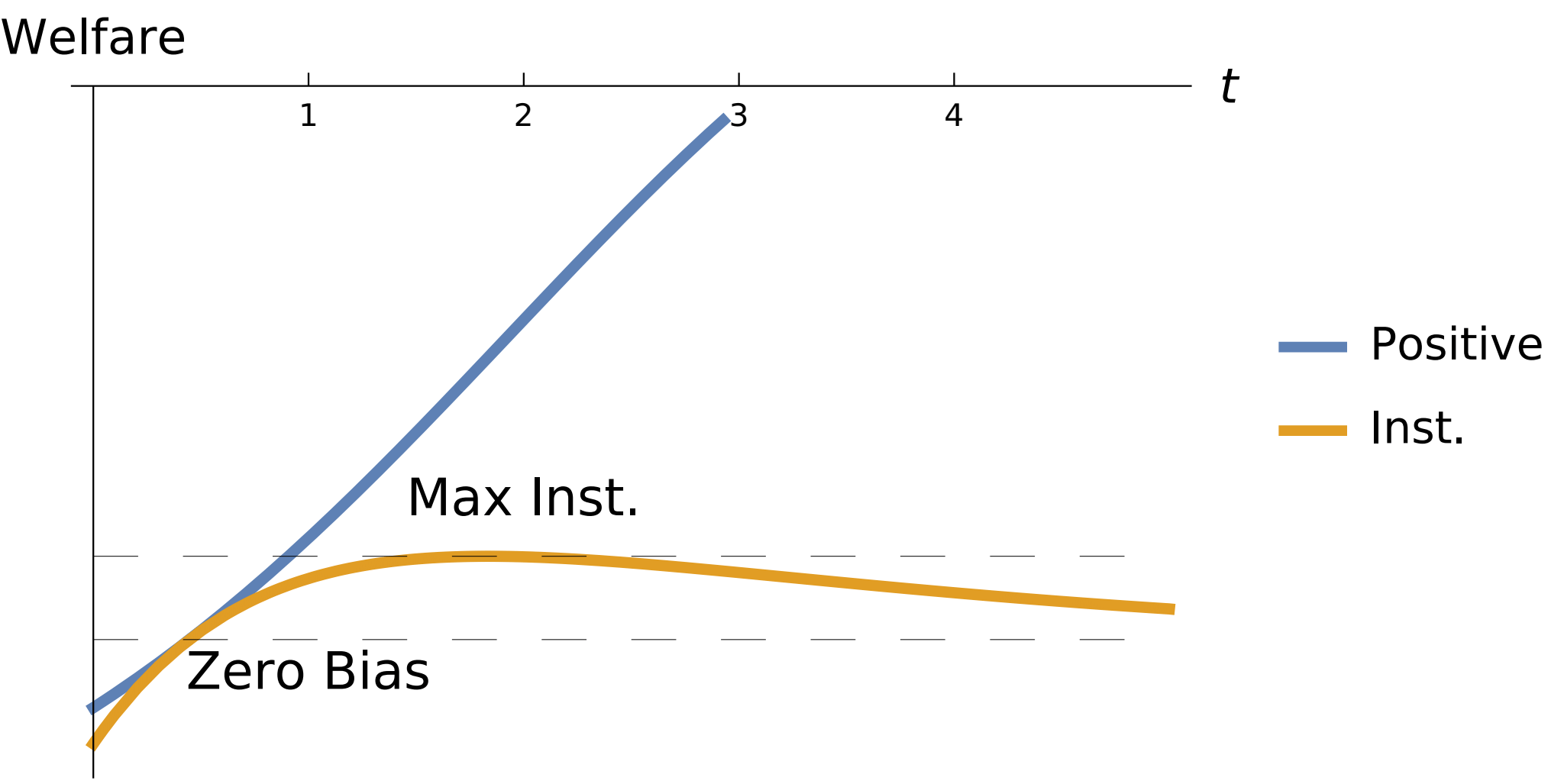}
   \caption{$\eta^{\star}\in(0,1)$ $\Rightarrow$ \text{Inst. Welfare has interior maximum, after the zero-bias time } $t^{\star}$}
   \label{fig:plot1}
\end{figure}

\begin{figure}
    \centering
    \includegraphics[scale=0.15]{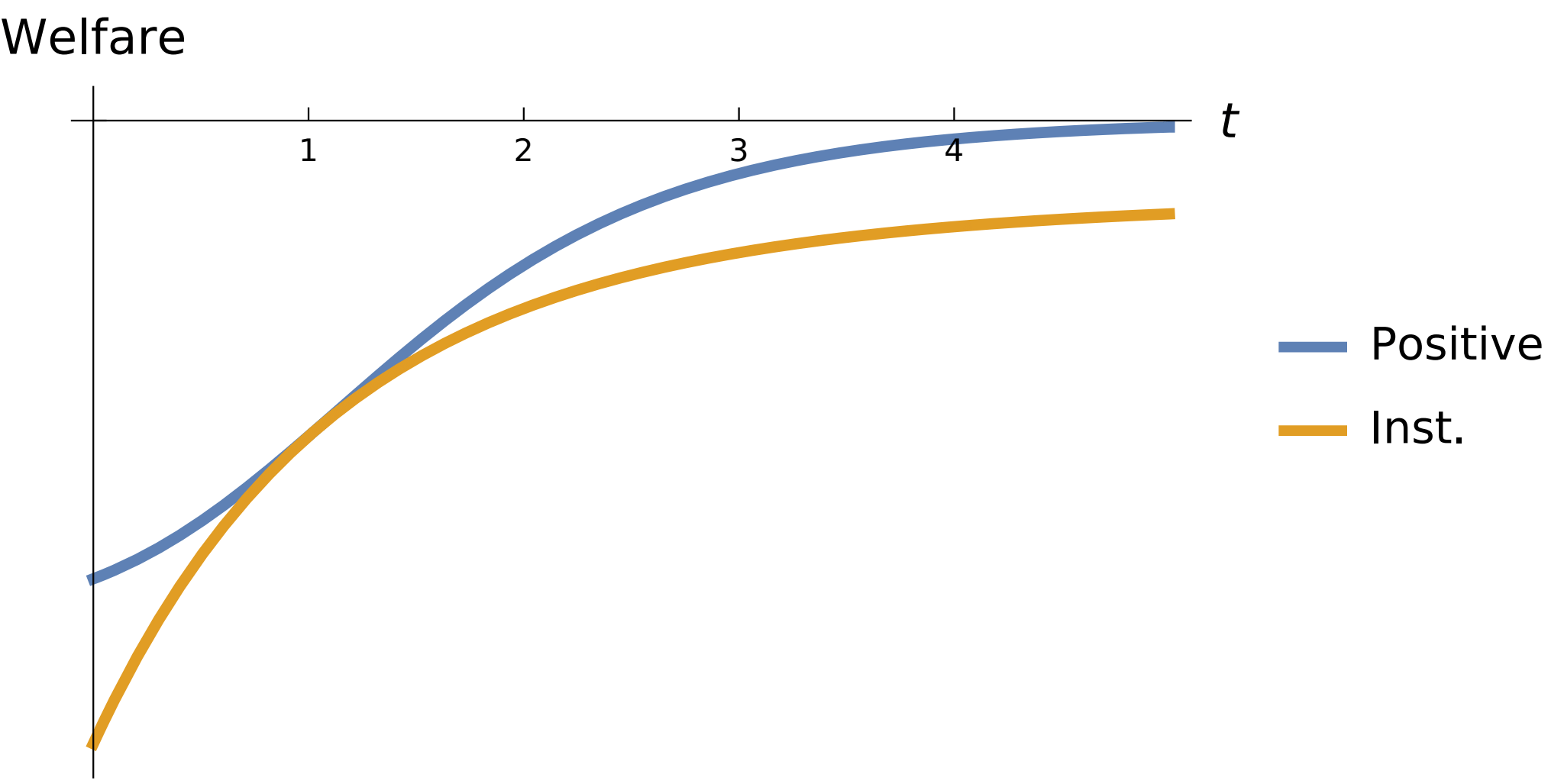}
   \caption{$\eta^{\star}=1$ $\Rightarrow$ \text{Inst. Welfare is increasing}}
   \label{fig:plot4}
\end{figure}

\begin{figure}[H]
    \centering
    \includegraphics[scale=0.16]{Plots/plot2.png}
   \caption{$\lambda$ \text{is a welfare measure.}}
   \label{fig:plot2}
\end{figure}

\subsection{Discussion of The Model Primitives}

\paragraph*{Implicit Assumptions.}

We have presented a relatively parsimonious model of voting while
undergoing a critical thinking process (from $S$tereotypes to $A$ware
citizens) of discovery of stable preferences. Its fundamental functioning
is easily explained. There is a nuisance parameter $p_{s}$ in the
sterotypes preferences that confounds the election outcome because
it adds a component that is orthogonal to the stable preference. As
more citizens become critical thinkers the election outcome is less
influenced by $p_{s}$ and the principal gets a better estimate of
the relevant parameter $p$. The share of critical thinkers increases
with time and with the intensity of the critical thinking process,
and this is the channel through which $\lambda$ impacts the efficiency
of elections. 

The two parameters $\beta,\xi$ are meant to capture the inherent
quality of the preference reported in the two stages of the critical
thinking process. High-$\beta$ environments represent situations
where, despite not realizing the ambivalent nature of the issue, stereotypes
get their stable preference right with high probability; it also makes
the independence assumption between $p$ and $p_{s}$ immaterial.
We think of $\xi$ instead as a reduced form parameter for a three-stages
critical thinking process in which agents first realize ambivalence,
and then discover the stable preference; $\xi$ is inversely proportional
to the length of this second transition. The flexibility added by
those two parameters does not alter the qualitative evolution of welfare
for a $P$ositive principal, but through their impact on the (evolution
of) bias they are consequential for an $I$nstitutional one, possibly
yielding to perplexing comparative statics. We have always maintained
an implicit symmetry assumption, since all structural parameters are
not allowed to depend on the stable preference $y$. A relaxation
of this assumption would require to model $\beta_{i}=\mathbb{P}\left[x_{S}=y|y=i\right],\xi_{i}=\mathbb{P}\left[x_{A}=y|y=i\right]$
with a different specification for the residual uncertainty in the
preference of stereotypes. Insofar as overconfidence can be interpreted
as individuals' resistance to critical thinking, evidence in \cite{ortoleva2015overconfidence} also questions the fact that the intensity $\lambda$
is independent of $y$: if stable preference predict cognitive traits
associated with critical thinking (or the impact of different storytelling
formats), then the $A$ware pool would be selected based on $y$,
which constitutes an additional source of bias. 

Extensions that allow these empirical regularities to manifest would
not alter the main message of our theoretical model: if there are
more $A$ware voters, polls contain more information about the distribution
of stable preferences. This is all that matters for a principal that
can ``filter out'' all systematic tendencies in voting --- including
the asymmetries in stereotype reporting and critical thinking transition
---, while a principal that cares about getting the election outcome
as close as possible to $p$ needs to trade off accuracy with election
bias. 

\paragraph*{Bias in Stereotypes and Critical Thinkers' Dilemma}

An immediate extension of our model is to allow for the presence of
bias in the stereotype pool, i.e. to let $\mu_{s}\neq\mu$, corresponding
to a situation where the principal knows that a specific opinion is
prevalent before individuals realize the ambivalent nature of an issue.
This possibility --- which seems compelling whenever one of the positions
is more prone to be defended by means of superficial arguments (nationalism)
--- means the $I$nstitutional principal additionally benefits from
increasing the intensity $\lambda$ (or simply ``letting time pass'')
as having a larger share of $A$ voters would mechanically remove
this type of systematic bias.\footnote{The evolution of welfare for the $P$ositive principal would instead
be unaffected by this extension, as she could ``clear out'' all
systematic noise in the poll. } Importantly, if this such bias became apparent to $A$ citizens too,
this might affect their voting rule. In particular, think about critical
thinkers who have not yet discovered their stable preference; their
problem is particularly interesting. Because they have ``lost''
their stereotype preference and not formed a stable one yet, they
most likely to abstain during an election and be more sensitive to
costly voting (an important margin, see Cantoni and Pons). If they
were aware of a systematic bias in stereotypes (which seems realistic,
since they just escaped that state) they might decide to use their
vote to ``compensate'' for such bias. 

Finally, by adding a penalty for waiting (discounting the utility
from taking an action later), we can use our setting to discuss the
optimal timing of elections; even the $P$ositive principal wouldn't
postpone her decision until $t=\infty$ where she would obtain a precise
estimate of $p$. Studying how the timing of the optimal election
varies with the intensity $\lambda$ (and other structural parameters)
amounts to analyzing the problem of a principal that controls the
type and duration of storytelling she wants to subject her agents
before administering a poll to maximize its accuracy. 
 
As for the impact of information, one implication is how different people react moving into and out of critical thinking, adding the amount of information, presenting a set of facts as bullet points, or re-elaborating those facts in a more structured piece. The challenge is that we do not observe people in critical thinking. Hence, how can we identify whether individuals are in critical thinking, and if so, how can we classify them as $S$ or $A$ throughout the different timing of exposure to storytelling formats?  We propose an experimental design and classification strategy to approximate the observation of such a critical thinking process.

\section{Experimental Setting}\label{experiment}

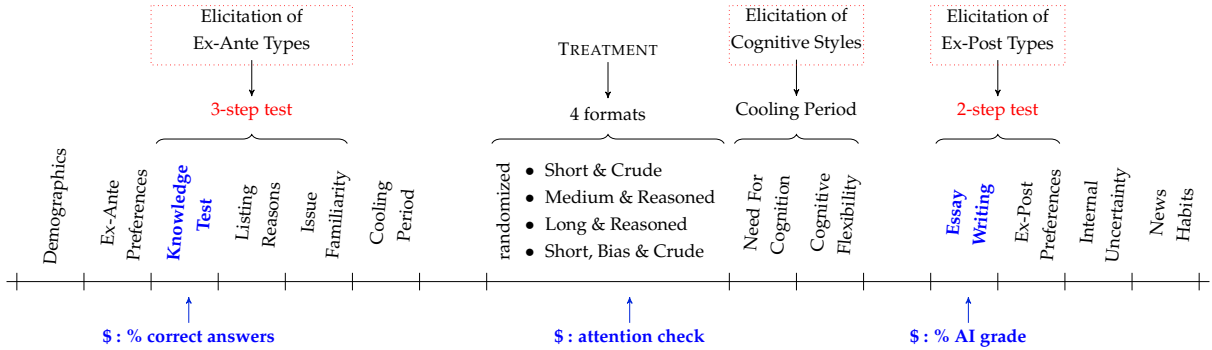
\begin{figure}[H]
\centering
\begin{adjustbox}{max totalsize={1\textwidth}{1\textheight},center}
\footnotesize
\begin{tikzpicture}[font=\footnotesize,every node/.style={align=center}]
		\draw[] (0,0)--(23.3,0);
		
		\foreach \i[count=\j] in {.2,1.5,...,10,14,15.3,...,24}
		\draw (\i,.17)--coordinate[pos=.5] (c\j) (\i,-.17);
		
		\node[above=3mm, align=left] (nc) at ($(c8)!.55!(c9)$) {
			$\bullet\,\,$ Short \& Crude\\
			$\bullet\,\,$ Medium \& Reasoned\\
			$\bullet\,\,$ Long \& Reasoned\\
			$\bullet\,\,$ Short, Bias \& Crude
		};
		
		\node[rotate=90,yshift=3mm] at (nc.west) {randomized};
		
	
		\draw [decorate,decoration = {brace, raise=5pt, amplitude=5pt}] ($(c3)+(0.1cm,2.5cm)$) -- node[above=5mm,align=center] (nn1) {\textcolor{red}{3-step test}} ($(c6)+(-0.1cm,2.5cm)$);
		
		\draw [decorate,decoration = {brace, raise=5pt, amplitude=5pt}] ($(c8)+(0.1cm,2.5cm)$) -- node[above=5mm,align=center] (nn2) {4  formats} ($(c9)+(-0.1cm,2.5cm)$);
		
		\draw [decorate,decoration = {brace, raise=5pt, amplitude=5pt}]  ($(c9)+(0.1cm,2.5cm)$) -- node[above=5mm,align=center] (nn3) {Cooling Period} ($(c11)+(-0.1cm,2.5cm)$);
		
		\draw [decorate,decoration = {brace, raise=5pt, amplitude=5pt}] ($(c12)+(0.1cm,2.5cm)$) -- node[above=5mm,align=center] (nn4) {\textcolor{red}{2-step test}} ($(c14)+(-0.1cm,2.5cm)$);
  
		\draw[stealth'-] ($(nn1)+(0,.3)$)-- node[pos=1,above] {Elicitation of\\Ex-Ante Types} ++(0,.7);
		\draw[stealth'-] ($(nn2)+(0,.3)$)-- node[pos=1,above] {\textsc{Treatment}} ++(0,.7);
		\draw[stealth'-] ($(nn3)+(0,.3)$)-- node[pos=1,above] {Elicitation of \\Cognitive Styles} ++(0,.7);
		\draw[stealth'-] ($(nn4)+(0,.3)$)-- node[pos=1,above] {Elicitation of \\Ex-Post Types} ++(0,.7);
		
		\coordinate (f1) at ($(c3)!.6!(c4)$);
		\draw[stealth'-, blue!70!teal] ($(f1)+(-.05,-.3)$)-- node[pos=1,below] {{\bf \textcolor{blue}{\$ : \% correct answers}}} ++(.0,-.5);
		\coordinate (f1) at ($(c8)!.6!(c9)$);
		\draw[stealth'-, blue!70!teal] ($(f1)+(-.05,-.3)$)-- node[pos=1,below] {{\bf \textcolor{blue}{\$ : attention check}}} ++(.0,-.5);
		\coordinate (f2) at ($(c12)!.6!(c13)$);
		\draw[stealth'-, blue!70!teal] ($(f2)+(-.05,-.3)$)-- node[pos=1,below] {{\bf \textcolor{blue}{\$ : \% AI grade}}} ++(0,-.5);
		\def\angle{85}
		\def\size{\footnotesize}
		\node[rotate=\angle, anchor=west, xshift=2mm, text width=2cm, font=\size] at ($(c1)!.5!(c2)$) {Demographics};
		\node[rotate=\angle, anchor=west, xshift=2mm, text width=2cm, font=\size] at ($(c2)!.5!(c3)$) {Ex-Ante Preferences};
		\node[rotate=\angle, anchor=west, xshift=2mm, text width=2cm, font=\size] at ($(c3)!.5!(c4)$) {\bf \textcolor{blue}{Knowledge Test}};
		\node[rotate=\angle, anchor=west, xshift=2mm, text width=2cm, font=\size] at ($(c4)!.5!(c5)$) {Listing Reasons};
		\node[rotate=\angle, anchor=west, xshift=2mm, text width=2cm, font=\size] at ($(c5)!.5!(c6)$) {Issue Familiarity};
		\node[rotate=\angle, anchor=west, xshift=2mm, text width=2cm, font=\size] at ($(c6)!.5!(c7)$) {Cooling Period};
		\node[rotate=\angle, anchor=west, xshift=2mm, text width=2cm, font=\size] at ($(c9)!.5!(c10)$) {Need For Cognition};
		\node[rotate=\angle, anchor=west, xshift=2mm, text width=2cm, font=\size] at ($(c10)!.5!(c11)$) {Cognitive Flexibility};
		\node[rotate=\angle, anchor=west, xshift=2mm, text width=2cm, font=\size] at ($(c12)!.5!(c13)$) {\bf \textcolor{blue}{Essay Writing}};
		\node[rotate=\angle, anchor=west, xshift=2mm, text width=2cm, font=\size] at ($(c13)!.5!(c14)$) {Ex-Post Preferences};
		\node[rotate=\angle, anchor=west, xshift=2mm, text width=2cm, font=\size] at ($(c14)!.5!(c15)$) {Internal Uncertainty};
		\node[rotate=\angle, anchor=west, xshift=2mm, text width=2cm, font=\size] at ($(c15)!.5!(c16)$) {News Habits};
		
		\draw[dotted, red] ([yshift=5.35cm]c3) rectangle ([yshift=4.2cm]c6);
  	\draw[dotted, red] ([yshift=5.35cm]c9) rectangle ([yshift=4.2cm]c11);
		\draw[dotted, red] ([yshift=5.35cm]c12) rectangle ([yshift=4.2cm]c14);
\end{tikzpicture}
\end{adjustbox}
\bigskip
\caption{Experiment Design} \label{fig:experiment}
\end{figure}

We have formalized an ambivalent environment in which  $\lambda$, interpreted as the speed of the critical thinking process, determines the efficiency of surveys and elections. We now empirically investigate whether different storytelling formats are associated with different  $\lambda$s. To do so, we estimate the $\lambda$ associated with different storytelling format treatments in an online incentivized experiment.\footnote{The Princeton Institutional Review Board approved the experiment. See the Appendix for the detailed Princeton IRB approval.} Figure \ref{fig:experiment} below provides an overview of the experimental design and its primary elicitations, which we will elaborate on in subsequent sections. 

In a nutshell, in our experiment, we expose subject participants to different storytelling formats and elicit their pre- and post-treatment stages in the critical thinking process associated with an ambivalent issue. We would then compare how the likelihood of transitioning from $S$ to $A$ varies across formats, testing whether they are statistically different. Throughout the experiment, we also collect data about participants' cognitive styles -- using standard measures from the psychological literature. This allows us to test whether the effectiveness of certain storytelling formats channels through identifiable cognitive traits. We used incentivized elicitations for the key individual variables —pre- and post-awareness states—and implemented anti-cheating policies and attention screeners to ensure optimal data collection quality. 

We gathered 900 participants from a representative US population using Prolific, a data collection platform increasingly favored by economists due to its high data quality. Following meticulous screening for attention, cheating, and quality, as outlined in the upcoming sections, our final sample size consisted of $N = 706$. Participants received a fixed payment of \$2 and a bonus payment of up to \$5, resulting in an average payment of approximately \$6.

\bigskip
\bigskip
\bigskip

\bigskip
\bigskip

\subsection{Classification of Critical Thinking States}\label{classstrategy}
We now present -- and describe the rationale for -- the two different strategies we employ to classify participants into the critical thinking states $\{S, A\}$ pre and post treatments. Table \ref{tab:globalclassstrat} summarizes both strategies. 

\paragraph{Pre-treatment classification strategy.} 

We use a three-pronged test to design our pre-treatment classification strategy as $\{S, A\}$. This test is based on the following heuristic conditions critical thinkers must satisfy: i) they must have basic knowledge of the issue at hand; ii) they must have thought about the issue before; iii) they must be aware that there exist both pros and cons for the issue. All i)-iii) characteristics are needed to be a critical thinker about an issue to avoid misclassification (as it could be by only using iii)). 

To elicit condition i), we rely on an assessment designed by Pew Research (\cite{vogels2019americans}) and launched on a representative US population, referred to as \textit{knowledge test} in Figure \ref{fig:experiment}. In our experiment, to pass the knowledge test, participants must score at least as high or above the score of the nationally-representative US population found by Pew Research.\footnote{It consists of 10 questions. See the Appendix for the wording details.} To elicit condition ii), we ask participants to report whether they have thought about the issue before coming to our experiment. To elicit condition iii), we ask them to provide evidence by providing two reasons that support their preference for the digital privacy issue and two that go against their preference. This task is instrumental in more accurately targeting the pre-treatment awareness state of individuals. We refer to this task as \textit{Listing reasons} is Figure \ref{fig:experiment}. If (and only if) subjects are already beyond their raw preference stage, we can provide a complete classification.
\\

We cannot rely on the same three-pronged tests to provide the post-treatment classification of participants in terms of $\{S,A\}$. Since the pre-treatment classification test includes condition iii) and our storytelling format treatments expose subjects to a series of pros and cons about the issue (see the next section), relying on the same condition here can misidentify critical thinking as memory effects. Indeed, subjects might not be critical thinkers, having accepted the issue as ambivalent by default, but happen to remember their list of pros and cons reported pre-treatment. Therefore, we need a different post-treatment classification strategy.

\paragraph{Post-treatment classification strategy.}
We classify participants' post-treatment critical thinking state as follows.  We require participants to write an incentivized essay discussing their preferences on the issue at hand. Subjects are instructed to present the issue and articulate their argumentative stance.\footnote{See Appendix to see the specific instructions to participants.} Their payment is based on the quality of their essay as measured by a Large Language Model (generative AI) powered software, Grammarly.  

Although Grammarly is efficient in assessing the overall quality of writing (at the time of our experiment, still powered by models similar to GPT-3), it lacks the capacity to capture the nuances of critical thinking, especially in terms of discerning whether the writer demonstrates an awareness of the ambivalence surrounding the issue. To address this limitation, we ask cognitive psychologists with Ph.D. degrees to provide a professional assessment of the essays. These experts are randomly assigned participants' essays and are asked to evaluate whether the essay reflects a state of awareness or not, assigning a pass or fail grade accordingly. While participants receive payment based on the AI's evaluation, our analysis focuses on the cognitive psychologists' assessment, with AI's scores serving as a robustness check (see Section \ref{subsec:robustness}). 

\bigskip

\begin{table}[H]
        \centering
        \footnotesize
        \addtolength{\tabcolsep}{5pt} 
        \begin{tabular}{lll}
        \toprule \toprule
        Treatment &\multicolumn{1}{c}{$ A $} & \multicolumn{1}{c}{$ S $} \\
        \midrule
                &  Knowledge Test Score $>$ $\tau_{KTS}$ \\
         \sc{before}    & Issue Familiarity = 1  & Else \\
                & Reasons List $>$ $\tau_{RL}$   \\
        \midrule
        \sc{after}    & Psychologists Grade = Pass & Else    \\
        \bottomrule \bottomrule
        \end{tabular}
        \caption{\sc Classification \textbf{strategy} before/after treatment}
        \label{tab:globalclassstrat}
\end{table}

\subsection{Measuring Cognitive Styles}\label{sec:cognitiveelicitations}

Since we are interested in explaining possible drivers of our results, through the experiment, we measure participants' cognitive styles and correlate those with the treatments effectiveness. We measure cognitive styles along three metrics. The first two are standard in the psychology literature: the Need for Cognitive Scale (NCS) and the Cognitive Flexibility Scale (CFS). 

NCS measures a participant's willingness to think deeply. It was proposed by neuroscientists, and cognitive psychologists \cite{cacioppo1982need} and have become a gold standard in cognitive psychology. It comprises a series of six questions that each receive a score between 1 and 5. We compare the aggregate score to the sample average to classify the participants into a high or low need for cognition.\footnote{No average for the US population is available for this score, despite being used widely across the social sciences and psychology. In addition, it was originally developed as a 34-question version, but the authors developed a shorter, more efficient 18-question version to elicit other psychological characteristics during the same laboratory session. Since then, it has been considered the benchmark scale widely used in cognitive and social sciences. An even shorter 6-question version has been tested and validated, allowing it to be implemented in a field survey experiment in which the participant's attention is even more scarce. We will use this later.} 

CFS measures an agent's ability to switch between thoughts and courses of action. It was proposed by cognitive psychologist \cite{martin1995new} and is a standard scale in cognitive psychology. It comprises a series of six questions that each receive a score between 1 and 6. We compare the aggregate score to the US population's average to classify participants into high or low cognitive flexibility.\footnote{The average is provided by the authors: 55.}

Finally, we use the AI generated score of an essay, unrelated to the core issue of our experiment, measuring individual's abilities to coherently present an argument.

\subsection{Description of Treatments}\label{sec:treatments}

Participants are randomly assigned to one of four treatments. These treatments contain the same  content (i.e., the same selection of facts about the digital issue, but they differ in the  semantic style and graphic design, as elaborated in the introduction) and last the same amount of time.

To recap, such formats range from the semantically crudest presentation of facts to the most refined presentation. The \textsc{twitter} treatment presents them most crudely through a ``tweet'' format. The \textsc{facebook} treatment uses the format of ``facebook posts.'' The \textsc{newspaper}  treatment presents them in the most refined way through ``newspaper articles.'' The \textsc{partisan Twitter} treatment uses only a partisan Twitter format (either only pros or cons)\footnote{In the appendix we detail and provide examples of each treatment.}.

Before the treatment starts, participants
 are explicitly informed that despite their high resemblance to real news, the tweets and Facebook posts are fake. At the end of the experiment, participants were debriefed and reminded that the tweets and Facebook posts were fake, following common practice in behavioral and experimental economics and as per our Institutional Review Board (IRB) approval.

\subsection{Incentive Mechanisms and Quality Screening}\label{sec:mechanismquality}

\paragraph{Incentive mechanism.} In the experiment, participants receive two types of payments. First, they receive a fixed reward of \$2 for completing the experiment fully by having answered the comprehension questions correctly, guaranteed. Second, they receive a bonus payment, at most \$6, as described below. 

Most of the participants' bonus payments (up to 5\$; participants' performance in the writing exercise, which captures their critical thinking process, determines their bonus). We ask the participants to write two short essays during this study that will be graded from 0 to 100 points by the artificial intelligence (AI)-powered software, Grammarly.\footnote{We, the authors, confirm to have neither professional ties nor a business contract with this company. See the Appendix for a summary of how this AI works.} We divide the bonus payment into two parts. 

The biggest part (from \$0 to \$5) is proportional to the weighted average score in the essay writing task; the second essay receives more weight (2/3) because it requires more writing (400 characters as opposed to 200 characters). The score can vary from 0 to 100 points, and the reward will be proportional to the score. If the participants get a score of 0, they win \$0. If they get a score of 50, then they win \$2.50. If they get a score of 100, then they win \$5. An essay that receives a low score from the AI can still earn a high score regarding critical thinking and awareness, despite the writer's difficulty with English. In the instructions to the psychologist graders, we define and exemplify what we mean by an ``ambivalent issue,'' ``realizing that the issue is ambivalent,'' and ``critical thinking''. We also run robust checks with philosophers.

To be eligible for the remaining bonus payment (up to \$1), participants must receive at least an average score of 50/100 in the essay exercise in addition to the bonus from the writing essay. Such a requirement ensures that participants do the exercise seriously; cheaters and agents that are inconsistent in their preferences are not eligible for this bonus payment. The participant's performance in the knowledge test determines this additional bonus. The test consists of 10 questions, and each participant receives \$0.10 for each question answered correctly.

\paragraph{Monitoring Algorithms for Cheating Behavior.}

We implement three attention screeners as it is standard in online experimental economics. The core of our experiment is for participants to write an original essay by themselves. We need the subjects to avoid accessing external information during the writing task. As such, we implement two algorithms to monitor cheating behavior. 

Before starting their experiment and on par with the IRB, we inform participants that they must not access external information during the experiment, particularly during the knowledge test and essay exercise. In addition, their essay must be original. Failing to do so would be considered ``cheating behavior.'' As such, they would be red-flagged and prevented from receiving anything other than the fixed payment. We excluded such participants from our data analysis.\footnote{We provide both algorithms as open source in our \href{https://github.com/brianjabarian/Critical-Thinking-and-Stable-Preferences}{GitHub}.}

The first algorithm tracks the number of times that the participants open a new tab on their computer during the essay exercise and how much time they spend on our essay writing web-page\footnote{For legal privacy purposes, we did not access the content of the opened tab but gathered only the following information: 'participant $i$ has opened a new tab during the essay, $n$ number of times, for such and such period $t$.}. The second algorithm checks whether the participants copy-paste external information by comparing the number of written characters and the number of keyboard clicks. If the number of keyboard clicks is strictly inferior to the number of written characters, it implies that the participants have copy-pasted external information. This second algorithm cannot distinguish between original external information\footnote{In the situation in which some participants had already written on the topic or a relevant topic and saved it on their computer before coming to the experiment.} and plagiarism. Therefore, we use a feature in the AI software to check for plagiarism after the participants have finished the experiment.

\section{Main Results}\label{mainresults}

\subsection{Storytelling formats Affect Critical Thinking}\label{sec:result}
We now test whether storytelling formats have a role in the critical thinking process of individuals. To this end, we compute for each treatment  $i=\{newspaper, twitter, facebook\}$, the frequency $\hat{\lambda}_i$ with which agents subject to  format $i$ transition from critical thinking state $S$ to critical thinking state $A$. Formally,
    \begin{align*}
        \hat{\lambda}_{i}= \frac{\# (S \rightarrow A)_{i}}{\# (S \rightarrow A)_{i} + \# (S \rightarrow S)_{i}}
    \end{align*}

We use the estimated intensities to perform a difference-in-means test of the null hypotheses $\lambda_{i} = \lambda_{j}$ for all possible combinations of  treatments $\{i, j\}$. 

Table \ref{tab:treatmentmajority} collects point estimates and confidence intervals.  From this table, we observe that the only significant difference is between \textit{Facebook} and \textit{Twitter}, where the former performs better in transitioning subjects from critical thinking state $S$ to $A$. Through this significant result, we establish that the  format impacts the critical thinking process. By being exposed to a different way of presenting the same basic information, individuals realize the ambivalent nature of the issue at hand differently. In section --, we perform robustness checks of this result (different thresholds, etc., metrics for success) to understand potential drivers of this effect.

\bigskip

 \begin{table}[H]
    \centering
    \footnotesize
    \addtolength{\tabcolsep}{5pt} 
    \begin{tabular}{lccc}
    \toprule \toprule
    Treatment &\multicolumn{1}{c}{\sc{newspaper}} & \multicolumn{1}{c}{\sc{twitter}}  & \multicolumn{1}{c}{\sc{facebook}}  \\
    \midrule
    \sc{newspaper} & $\cdot$ & 1.332 & -0.865      \\
    &  & (0.054) & (0.054) \\
    {\sc{twitter}}    &  $\cdot$ & $\cdot$ & -2.249\sym{**}        \\
    &  &  & (0.053) \\
    \sc{facebook} & $\cdot$ & $\cdot$ & $\cdot$ \\ \bottomrule \bottomrule
    \multicolumn{3}{l}{\footnotesize Standard errors in parentheses}\\
    \multicolumn{3}{l}{\footnotesize \sym{*} \(p<0.1\), \sym{**} \(p<0.05\), \sym{***} \(p<0.01\)}\\
    \end{tabular}
    \caption{\sc \textit{t-ratio} difference-in-means}
    \label{tab:treatmentmajority}
\end{table}
\bigskip

A possible explanation for the observed difference in the impact of the storytelling formats on critical thinking, as highlighted in Table \ref{tab:treatmentmajority}, is that the treatment \textsc{\bfseries twitter} may rely on a format that is too simplistic or naive to nudge users towards critical thinking effectively. This explanation can be further elaborated as follows.

First, while Twitter imposes a character limit on its content, forcing users to use concise language and simplifying complex ideas, Facebook allows for longer and more detailed post\footnote{This experiment was designed and launched before Musk Twitter's area, which led to the increase of tweets lengths for Blue Twitter users, which now can be considered as our Facebook treatment.}. This difference in content structure could impact how individuals process information and engage in critical thinking.

Second, the fast-paced nature of Twitter feeds, and the emphasis on real-time information sharing could discourage users from pausing, reflecting, and analyzing the content they consume. This constant influx of new information might contribute to a shallower engagement with the material, reducing the likelihood of critical thinking.

Third, Twitter's focus on short, attention-grabbing headlines and sound bites may encourage users to form quick opinions based on surface-level information rather than delving deeper into the nuances of an issue. This aspect of the platform's design might hinder the development of well-informed perspectives and critical thinking.

Fourth, the prevalence of echo chambers on Twitter, where users primarily follow and interact with those who share their views, could further contribute to the observed limitations of the Twitter format in promoting critical thinking. This selective exposure to information might reinforce pre-existing beliefs and discourage users from challenging their assumptions.

Fifth, another factor to consider is the nature of user engagement on these platforms. Facebook is known for fostering more personal connections and allowing in-depth conversations, while Twitter primarily emphasizes short, quick information exchanges. This contrast in user engagement could contribute to the observed difference in the effectiveness of the storytelling formats on critical thinking.

Finally, the role of media consumption habits might be influential in explaining the difference in critical thinking outcomes. Users of Facebook may be more inclined to read longer posts and engage in reflective thinking, whereas Twitter users might be more accustomed to quickly skimming through bite-sized information. As a result, individuals' media consumption habits could shape their receptiveness to the storytelling formats on these platforms, ultimately affecting their critical thinking process.

\subsection{Heterogeneity in Cognitive Styles}\label{cognitiveresults}
We explore whether the cognitive traits we elicited explain the differential effect by conducting a split-sample difference in means. We test whether $\lambda_i=\lambda_j$ by partitioning our sample into high or low individuals in our cognitive metrics. The idea is that a more in-depth  (like the journal article) might be more effective for individuals more prone to think deeply.The efficacy of the \textit{Facebook} treatment was driven by its differential impact on \textit{High Need for Cognition} agents.

\begin{table}[H]
        \centering
        \footnotesize
        \begin{tabular}{lccc}
        \toprule \toprule
        Treatment &\multicolumn{1}{c}{\sc{NP}} & \multicolumn{1}{c}{\sc{twitter}}  & \multicolumn{1}{c}{\sc{facebook}}  \\
        \midrule
        \sc{newspaper} & $\cdot$ & 0.764 & -2.238\sym{*}      \\
        &  & (0.070) & (0.079) \\
        \sc{twitter}   &  $\cdot$ & $\cdot$ & -3.087\sym{**}        \\
        &  &  & (0.075) \\
        \sc{facebook}   & $\cdot$ & $\cdot$ & $\cdot$           \\
        \bottomrule \bottomrule
        \multicolumn{3}{l}{\footnotesize Standard errors in parentheses}\\
        \multicolumn{3}{l}{\footnotesize \sym{*} \(p<0.05\), \sym{**} \(p<0.01\), \sym{***} \(p<0.001\)}\\
        \end{tabular}
         \caption{\sc \textit{t-ratio} for \textit{High Need for Cognition.}}
        \label{tab:hnfcmajority}
\end{table}

These results suggest that subjects who are the most affected by the storytelling format are those who exhibit a high need for cognition. For them, treatment \textsc{\bfseries facebook} seems to provide the right format to maximally capture their attention to present an issue so that it successfully nudges them into performing the critical thinking process. 

One possible explanation for the results observed in Table \ref{tab:hnfcmajority} could be rooted in the characteristics of individuals with a High Need for Cognition. These individuals typically exhibit a greater tendency to engage in effortful cognitive activities and prefer more complex information processing (\cite{cacioppo1982need}). Consequently, the Facebook format might provide a more stimulating environment for critical thinking by offering a richer and more nuanced presentation of information than the Twitter format.

Furthermore, it has been suggested that individuals with a higher need for cognition are more likely to seek out, attend to, and remember information consistent with their attitudes and beliefs (Hass \& Linder, 1981). As a result, the Facebook format could be more effective in capturing their attention and motivating them to evaluate the content critically. This might explain why the Facebook treatment significantly impacts transitioning subjects from critical thinking state S to A among those with a High Need for Cognition.

Future research could explore the specific features of the Facebook format that contribute to its efficacy in promoting critical thinking among individuals with a high need for cognition. For instance, it would be interesting to investigate the role of multimedia elements, interactivity, and the integration of diverse information sources in fostering an environment conducive to critical thinking.

\subsection{Robustness Analysis}\label{subsec:robustness}

We address two potential challenges to ensure the robustness of our findings -- threshold sensitivity and writing similarity checks -- that we present now.

\paragraph{Threshold sensitivity.} Our conclusions should remain consistent regardless of 
the specific values of the threshold used in characteristics i) and iii) of the three-pronged test we use to classify participants pre-treatment. Recall that i) refers to the digital knowledge test and iii) refers to the reasons listing exercise. 

Regarding i), in the baseline, we require participants to score at least 7 correct answers out of 10 questions. Comparing to the original setting provided by Pew Research, our threshold is much more demanding. The quiz by Pew Research was launched on a large US representative sample of 4,272 adults living in the United States. The median number of correct answers was four. Only 20\% of adults answered seven or more questions correctly, and just 2\% got all 10 questions correct. Despite this difference, still, we are interested to check whether our treatment effectiveness depends on scoring higher or lower than scoring 7 out of 10. 

Regarding iii), in the baseline, we require participants to be capable of listing at least one reason for one side (pro or con) and two reasons for the other side (pro or con). We are interested to check whether our treatment effectiveness depend on the capacity of participants to list more than one reason for each side. 

\paragraph{Writing similarity.} Our findings should not be influenced by the similarity in length between the essay task and any specific treatment, particularly the Facebook treatment. By comparing outcomes across different essay lengths or imposing length constraints, we can verify that the observed effects are not artifacts of such similarities, ensuring the robustness of our results.

Overall, our robustness analysis confirm that the treatment effectiveness does not depend neither on the threshold sensitivity test nor on the writing similarity test.\footnote{We provide the analysis in the appendix.}

\section{Conclusion}\label{conclusion}
In this paper, we built a simple but flexible model to measure the gain in election efficiency by becoming aware that the issue is ambivalent ($A$). We experimented and determined that the format in which news is presented affects a person’s transition into $A$. This effect is driven by individuals with a high need for cognition (the flexibility scale is insignificant). Realizing the ambivalent nature of an issue is an essential step in discovering one's stable preference since it improves the ``quality'' of one’s preference from raw to stable. As such, critical thinking is \emph{also} good for the efficiency of elections. 

What is broadly referred to as a storytelling format (e.g., newspapers, television, social media, social echo chambers) might impact the probability of realizing ambivalence. Beyond ``informing'' and ``persuading,'' it also affects an individual's critical thinking process. Additionally, the format in which news is presented---many short messages vs. more coherent but greedy attention discourse---matters. In particular, unexplored (by us) physiological drivers were correlated with standard metrics of cognition/flexibility.

Stable preferences $y$ are \emph{not observable}, and the model is not (fully)
 identified. We rely on a reduced form for an identified model with three cognitive stages $S\to A\to T$ and a final transition to a stable preference (resolving
awareness) with qualitatively similar results. First, $=A$ realized the issue was ambivalent but still did not find our $y$. Second, how did voters in $A$ vote? Strategic voting in the presence of stereotype bias, ``I still have not resolved my awareness about {[}topic{]}, but I see much prejudice in favor of position $0$, so I vote $1$
 to compensate.'' Additionally, stable preference $y$ is independent of other individual types ($\beta,\xi_{A},\lambda\dots$). Prejudices often coincide with a stable preference if the latter is $1$. $\beta_{1}>\beta_{0}$. Prejudices might correlate with the likelihood of becoming aware (\cite{ortoleva2015overconfidence}).

Regarding the internal validity of our experiment, we recognize that classifying individuals’ critical thinking states is inherently challenging. We devised different classification rules for pre- and post-treatment to avoid mistaking memory for critical thinking. Second, we used a noisy measure to look at the \textit{difference across treatments}. Regarding the external validity of our experiment, readers should refrain from interpreting our experiment as a comparison of social media, concluding that ``Facebook is better'' but rather ``the
 format matters.'' In this interpretation, the whole class of social media becomes a storytelling format: one is exposed to a greater number of views, but they are possibly superficial. Does it help to become aware of the issue's ambivalence relative to one’s life experience or the in-depth study of a topic (more personal and reasoned but time-consuming and unlikely to occur)?

\bibliography{bibnormuncer}

\begin{thebibliography}{23}
\providecommand{\enquote}[1]{``#1''}
\providecommand{\natexlab}[1]{#1}
\providecommand{\url}[1]{\texttt{#1}}
\providecommand{\urlprefix}{URL }
\providecommand{\bibAnnoteFile}[1]{%
  \IfFileExists{#1}{\begin{quotation}\noindent\textsc{Key:} #1\\
  \textsc{Annotation:}\ \input{#1}\end{quotation}}{}}
\providecommand{\bibAnnote}[2]{%
  \begin{quotation}\noindent\textsc{Key:} #1\\
  \textsc{Annotation:}\ #2\end{quotation}}

\bibitem[{Aragones et~al.(2005)Aragones, Gilboa, Postlewaite, and
  Schmeidler}]{aragones2005fact}
Aragones, Enriqueta, Itzhak Gilboa, Andrew Postlewaite, and David Schmeidler
  (2005), \enquote{Fact-free learning.} \emph{American Economic Review}, 95,
  1355--1368.
\bibAnnoteFile{aragones2005fact}

\bibitem[{B{\'e}nabou*(2015)}]{benabou2015economics}
B{\'e}nabou*, Roland (2015), \enquote{The economics of motivated beliefs.}
  \emph{Revue d'{\'e}conomie politique}, 665--685.
\bibAnnoteFile{benabou2015economics}

\bibitem[{Bernheim et~al.(2021)Bernheim, Braghieri, Mart{\'\i}nez-Marquina, and
  Zuckerman}]{bernheim2021theory}
Bernheim, B~Douglas, Luca Braghieri, Alejandro Mart{\'\i}nez-Marquina, and
  David Zuckerman (2021), \enquote{A theory of chosen preferences.}
  \emph{American Economic Review}, 111, 720--54.
\bibAnnoteFile{bernheim2021theory}

\bibitem[{Cacioppo and Petty(1982)}]{cacioppo1982need}
Cacioppo, John~T and Richard~E Petty (1982), \enquote{The need for cognition.}
  \emph{Journal of personality and social psychology}, 42, 116.
\bibAnnoteFile{cacioppo1982need}

\bibitem[{Conway~III et~al.(2012)Conway~III, Gornick, Burfeind, Mandella,
  Kuenzli, Houck, and Fullerton}]{conway2012does}
Conway~III, Lucian~Gideon, Laura~Janelle Gornick, Chelsea Burfeind, Paul
  Mandella, Andrea Kuenzli, Shannon~C Houck, and Deven~Theresa Fullerton
  (2012), \enquote{Does complex or simple rhetoric win elections? an
  integrative complexity analysis of us presidential campaigns.}
  \emph{Political Psychology}, 33, 599--618.
\bibAnnoteFile{conway2012does}

\bibitem[{Falck et~al.(2014)Falck, Gold, and Heblich}]{falck2014}
Falck, Oliver, Robert Gold, and Stephan Heblich (2014), \enquote{E-lections:
  Voting behavior and the internet.} \emph{American Economic Review}, 104,
  2238--2265.
\bibAnnoteFile{falck2014}

\bibitem[{Feddersen and Pesendorfer(1997)}]{feddersen1997voting}
Feddersen, Timothy and Wolfgang Pesendorfer (1997), \enquote{Voting behavior
  and information aggregation in elections with private information.}
  \emph{Econometrica: Journal of the Econometric Society}, 1029--1058.
\bibAnnoteFile{feddersen1997voting}

\bibitem[{Gorodnichenko et~al.(2021)Gorodnichenko, Pham, and
  Talavera}]{gorodnichenko2021}
Gorodnichenko, Yuriy, Tho Pham, and Oleksander Talavera (2021), \enquote{Social
  media, sentiment and public opinions: Evidence from \#brexit and
  \#uselection.} \emph{European Economic Review}, 136.
\bibAnnoteFile{gorodnichenko2021}

\bibitem[{Gul and Pesendorfer(2009)}]{gul2009partisan}
Gul, Faruk and Wolfgang Pesendorfer (2009), \enquote{Partisan politics and
  election failure with ignorant voters.} \emph{Journal of Economic Theory},
  144, 146--174.
\bibAnnoteFile{gul2009partisan}

\bibitem[{Halpern(2013)}]{halpern2013thought}
Halpern, Diane~F (2013), \emph{Thought and knowledge: An introduction to
  critical thinking}. Psychology Press.
\bibAnnoteFile{halpern2013thought}

\bibitem[{Jordan et~al.(2019)Jordan, Sterling, Pennebaker, and
  Boyd}]{jordan2019examining}
Jordan, Kayla~N, Joanna Sterling, James~W Pennebaker, and Ryan~L Boyd (2019),
  \enquote{Examining long-term trends in politics and culture through language
  of political leaders and cultural institutions.} \emph{Proceedings of the
  National Academy of Sciences}, 116, 3476--3481.
\bibAnnoteFile{jordan2019examining}

\bibitem[{Kahneman(2011)}]{kahneman2011thinking}
Kahneman, Daniel (2011), \emph{Thinking, fast and slow}. Macmillan.
\bibAnnoteFile{kahneman2011thinking}

\bibitem[{Kaplan(1972)}]{kaplan1972}
Kaplan, Kalman~J (1972), \enquote{On the ambivalence-indifference problem in
  attitude theory and measurement: A suggested modification of the semantic
  differential technique.} \emph{Psychological Bulletin}, 77, 361--372.
\bibAnnoteFile{kaplan1972}

\bibitem[{Kim and Fey(2007)}]{kim2007swing}
Kim, Jaehoon and Mark Fey (2007), \enquote{The swing voter's curse with
  adversarial preferences.} \emph{Journal of Economic Theory}, 135, 236--252.
\bibAnnoteFile{kim2007swing}

\bibitem[{Kunda(1990)}]{kunda1990case}
Kunda, Ziva (1990), \enquote{The case for motivated reasoning.}
  \emph{Psychological bulletin}, 108, 480.
\bibAnnoteFile{kunda1990case}

\bibitem[{List(2022)}]{list2022enhancing}
List, John~A (2022), \enquote{Enhancing critical thinking skill formation:
  Getting fast thinkers to slow down.} \emph{The Journal of economic
  educaTion}, 53, 100--108.
\bibAnnoteFile{list2022enhancing}

\bibitem[{Martin and Rubin(1995)}]{martin1995new}
Martin, Matthew~M and Rebecca~B Rubin (1995), \enquote{A new measure of
  cognitive flexibility.} \emph{Psychological reports}, 76, 623--626.
\bibAnnoteFile{martin1995new}

\bibitem[{Millner(2020)}]{millner2020}
Millner, Anthony (2020), \enquote{Nondogmatic social discounting.}
  \emph{American Economic Review}, 110, 760--775.
\bibAnnoteFile{millner2020}

\bibitem[{Munir(2018)}]{munir2018social}
Munir, Saba (2018), \enquote{Social media and shaping voting behavior of youth:
  The scottish referendum 2014 case.} \emph{The Journal of Social Media in
  Society}, 7, 253--279.
\bibAnnoteFile{munir2018social}

\bibitem[{Ortoleva and Snowberg(2015)}]{ortoleva2015overconfidence}
Ortoleva, Pietro and Erik Snowberg (2015), \enquote{Overconfidence in political
  behavior.} \emph{American Economic Review}, 105, 504--35.
\bibAnnoteFile{ortoleva2015overconfidence}

\bibitem[{Tetlock(1981)}]{tetlock1981pre}
Tetlock, Philip~E (1981), \enquote{Pre-to postelection shifts in presidential
  rhetoric: Impression management or cognitive adjustment.} \emph{Journal of
  Personality and Social Psychology}, 41, 207.
\bibAnnoteFile{tetlock1981pre}

\bibitem[{Thoemmes and Conway~III(2007)}]{thoemmes2007integrative}
Thoemmes, Felix~J and Lucian~Gideon Conway~III (2007), \enquote{Integrative
  complexity of 41 us presidents.} \emph{Political Psychology}, 28, 193--226.
\bibAnnoteFile{thoemmes2007integrative}

\bibitem[{Vogels and Anderson(2019)}]{vogels2019americans}
Vogels, Emily~A and Monica Anderson (2019), \enquote{Americans and digital
  knowledge.}
\bibAnnoteFile{vogels2019americans}

\end{thebibliography}

\newpage

\appendix

\appendixpage
\startcontents[sections]
\printcontents[sections]{l}{1}{\setcounter{tocdepth}{2}}

\section{Proofs of The Main Model}

\subsection{Preliminary Results on $\hat{p}$ and alike}

The following three steps explicitly show how to analyse the evolution of the cognitive state process over time for each agent and how this relates to the parameters of the model.
\\

1) $\mu_{S}=\exp\left\{ -\lambda_{1}t\right\} $ and $\mu_{C}=1-\mu_{S}$
represent the masses. $\lambda_{1}$ represents the intensity with
which agents pass from the cognitive state $S$ to the cognitive state
$A$ over time. Moreover, we define the unknown parameter $\bar{p}$
as function of $\mu$ and $p$

\begin{align*}
\overline{p}\left(\mu,p\right) & =\mu_{S}\left(\mathbb{E}\left[x_{S}\left|p\right.\right]\right)+\mu_{C}\left(\mathbb{E}\left[x_{C}\left|p\right.\right]\right)\\
 & =\mu_{S}\left(\beta p_{S}+\left(1-\beta\right)p\right)+\mu_{C}\left(\xi_{C}p+\left(1-\xi_{C}\right)\left(1-p\right)\right)\\
 & =\mu_{S}\left(\beta p_{S}+\left(1-\beta\right)p\right)+\mu_{C}\left(1-p-\xi_{C}\left(1-2p\right)\right)
\end{align*}
\\
Thus

\begin{equation}
\overline{p}\left(\mu,p\right)=\mu_{S}\left(\beta p_{S}+\left(1-\beta\right)p\right)+\mu_{C}\left(1-p-\xi_{C}\left(1-2p\right)\right)
\end{equation}

From which we can derive the expression for $p$ as a function of $p_{S}$

\begin{align*}
\bar{p} & =\mu_{S}\beta p_{S}+\mu_{S}\left(1-\beta\right)p+\mu_{C}-\mu_{C}p-\mu_{C}\xi_{C}+2\mu_{C}p\xi_{C}\\
\bar{p} & =\mu_{S}\beta p_{S}+\mu_{C}-\mu_{C}\xi_{C}+p\left[\mu_{S}\left(1-\beta\right)-\mu_{C}\left(1-2\xi_{C}\right)\right]\\
 & =\mu_{S}\beta p_{S}+\mu_{C}-\mu_{C}\xi_{C}+p\left[\mu_{S}\left(1-\beta\right)-\mu_{C}\left(1-2\xi_{C}\right)\right]
\end{align*}
Thus $p$ is defined as \\
\begin{equation}
p=\frac{\bar{p}-\mu_{S}\beta p_{S}-\mu_{C}\left(1-\xi_{C}\right)}{\mu_{S}\left(1-\beta\right)-\mu_{C}\left(1-2\xi_{C}\right)}
\end{equation}
\\
Finally, we can define the parameter $\hat{p}$ that is defined as
the expectation of $p$ conditioning on $\bar{p}$

\begin{equation}
\hat{p} \& =\frac{\bar{p}-\left[\mu_{S}\beta\mathbb{E}\left[p_{S}|\bar{p}\right]+\mu_{C}\left(1-\xi_{C}\right)\right]}{\mu_{S}\left(1-\beta\right)+\mu_{C}\left(2\xi_{C}-1\right)}
\end{equation}

Thus $\bar{p}$ is defined as

\begin{equation}
\hat{p}=\mathbb{E}\left[p|\alpha_{1}p+\alpha_{2}p_{S}=\bar{p}\right]
\end{equation}

\begin{align*}
\hat{p} & =\frac{\beta^{2}\mu_{S}^{2}\mu_{X}\sigma_{Y}^{2}+\sigma_{X}^{2}\left(1-\bar{p}-\xi_{C}+\mu_{S}\left(-1+\beta\mu_{Y}+\xi_{C}\right)\right)\left(1-2\xi_{C}+\mu_{S}\left(-2+\beta+2\xi_{C}\right)\right)}{\beta^{2}\mu_{S}^{2}\sigma_{Y}^{2}+\sigma_{X}^{2}\left(1-2\xi_{C}+\mu_{S}\left(-2+\beta+2\xi_{C}\right)\right)^{2}}\\
 & =\frac{\beta^{2}\mu_{S}^{2}\mu_{X}\sigma_{Y}^{2}+\sigma_{X}^{2}\left(\bar{p}-\left[\mu_{S}\beta\mu_{y}+\left(1-\mu_{S}\right)\left(1-\xi_{C}\right)\right]\right)\left(1-2\xi_{C}+\mu_{S}\left(-2+\beta+2\xi_{C}\right)\right)}{\beta^{2}\mu_{S}^{2}\sigma_{Y}^{2}+\sigma_{X}^{2}\left(1-2\xi_{C}+\mu_{S}\left(-2+\beta+2\xi_{C}\right)\right)^{2}}
\end{align*}
\\

2) It is worth noting that the NWF can be expressed as the sum of the
PWF and a biased term due to the elections. Indeed,\\

\begin{align*}
NWF & =-\mathbb{E}\left[(p-\bar{p})^{2}\right]=-\mathbb{E}\left[\left(p-\hat{p}+\hat{p}-\bar{p}\right)^{2}\right]\\
 & =-\mathbb{E}\left[(p-\hat{p})^{2}+2(p-\hat{p})(\hat{p}-\bar{p})+(\hat{p}-\bar{p})^{2}\right]\\
 & =-\left[\mathbb{E}\left[(p-\hat{p})^{2}\right]+2\mathbb{E}\left[(p-\hat{p})(\hat{p}-\bar{p})\right]+\mathbb{E}\left[(\hat{p}-\bar{p})^{2}\right]\right]\\
 & =-\left[\mathbb{E}\left[(p-\hat{p})^{2}\right]+2(\hat{p}-\bar{p})\underbrace{\mathbb{E}\left[(p-\hat{p})\right]}_{0}+\mathbb{E}\left[(\hat{p}-\bar{p})^{2}\right]\right]\\
 & =-\left[\underbrace{\mathbb{E}\left[(p-\hat{p})^{2}\right]}_{\text{Precision of elections}}+\underbrace{\mathbb{E}\left[(\hat{p}-\bar{p})^{2}\right]}_{\text{Bias of elections}}\right]
\end{align*}
 
Thus it can be rewritten as

\[
NWF=PWF+Bias
\]
\\
At this stage, we define the two welfare functions given the distributions
of the parameters

3)What the theoretical analysis wants to show is the evolution of the
welfare functions over time and the main differences between the evolution
of the PWF and the NWF. In particular, in order to study the evolution,
we take the first derivative of the two functions with respect to
$\mu_{S}$. It is necessary and sufficient to show the sign of this
derivative in order to have an all rounded understanding of the evolution
of the two functions. Indeed, $\mu_{S}$ as defined above depends
negatively on $t$ and $\lambda_{1}$. Hence, once we define the relation
between the functions and $\mu_{S},$ we immediately get to know the
relation between the functions and the time/lambda. Thus let's start
from showing the behavior of the PWF\\

\begin{align*}
\frac{\partial PWF}{\partial\mu_{S}} & =\\
\frac{2\beta^{2}\mu_{S}\sigma^{2}\left(-1+2\xi_{C}\right)\left[1-2\xi_{C}+\mu_{S}\left(-2+\beta+2\xi_{C}\right)\right]}{\left\{ 2\mu_{S}^{2}\left[\beta^{2}+2\beta\left(-1+\xi_{C}\right)+2\left(-1+\xi_{C}\right)^{2}\right]+\left(1-2\xi_{C}\right)^{2}-2\mu_{S}\left(-1+2\xi_{C}\right)\left(-2+\beta+2\xi_{C}\right)\right\} ^{2}}\\
 & \propto1-2\xi_{C}+\mu_{S}\left(\beta-2\left(1-\xi_{C}\right)\right)
\end{align*}
\\
Since almost everything is bigger or equal than $0$, if we want to
study the sign of the above formula, then we just have to analyse
the sign of the following term\\

\begin{align*}
1-2\xi_{C}+\mu_{S}\left(-2+\beta+2\xi_{C}\right) & <0\\
0<\mu_{S} & <\underbrace{\frac{2\xi_{C}-1}{-2+\beta+2\xi_{C}}}_{\geq1?}
\end{align*}

\begin{prop}
$W_{P}$ is increasing in $t$ and $\lambda$ if 

\[
\frac{2\xi_{C}-1}{-2+\beta+2\xi_{C}}>1\iff1>\beta
\]
\end{prop}

Let's study the right hand side of the inequality

\begin{align*}
2\xi_{C}-1 & \geq-2+\beta+2\xi_{C}\\
\beta & \leq1
\end{align*}

Therefore, we can conclude that PWF is decreasing in $\mu_{S}$ for
each time $t$, because

\[
0<\mu_{S}<\frac{2\xi_{C}-1}{-2+\beta+2\xi_{C}},\;\;\;\forall\mu_{S}\in[0,1]
\]
\\
In other words, the PWF is an increasing function of both $t$ and
$\lambda_{1}$.\\

\subsection{Proof of Proposition \ref{prop:Main-Result}}

\begin{proof}
Notice preliminary that using the chain rule the following result is valid for both welfare functions

\[
\frac{\text{d}W}{\text{d}\lambda}=\frac{\text{d}W}{\text{d}\eta}\cdot\underbrace{\frac{\text{d}\eta}{\text{d}\lambda}}_{<0}\implies\frac{\text{d}W}{\text{d}\lambda}\propto-\frac{\text{d}W}{\text{d}\eta}
\]
 and therefore welfare moves in $\lambda$ (and $t$) contrary to
how it moves in the share of stereotypes. First, for positive welfare,
$\frac{\text{d}W}{\text{d}\eta}$ is always negative for the following
computations

\begin{equation*}
\begin{split}
\frac{\partial PWF}{\partial\mu_{S}}=\\
\frac{2\beta^{2}\mu_{S}\sigma^{2}\left(-1+2\xi_{C}\right)\left[1-2\xi_{C}+\mu_{S}\left(-2+\beta+2\xi_{C}\right)\right]}{\left\{ 2\mu_{S}^{2}\left[\beta^{2}+2\beta\left(-1+\xi_{C}\right)+2\left(-1+\xi_{C}\right)^{2}\right]+\left(1-2\xi_{C}\right)^{2}-2\mu_{S}\left(-1+2\xi_{C}\right)\left(-2+\beta+2\xi_{C}\right)\right\} ^{2}} \\ \propto1-2\xi_{C}+\mu_{S}\left(\beta-2\left(1-\xi_{C}\right)\right)
\end{split}
\end{equation*}
Since almost everything is bigger or equal than $0$, if we want to
study the sign of the above formula, then we just have to analyse
the sign of the following term
\begin{align*}
1-2\xi_{C}+\mu_{S}\left(-2+\beta+2\xi_{C}\right) & <0\\
0<\mu_{S} & <\underbrace{\frac{2\xi_{C}-1}{-2+\beta+2\xi_{C}}}_{\geq1?}
\end{align*}
Let's study the right hand side of the inequality
\begin{align*}
2\xi_{C}-1 & \geq-2+\beta+2\xi_{C}\\
\beta & \leq1
\end{align*}
Therefore, we can conclude that PWF is decreasing in $\mu_{S}$ for
each time $t$, because
\[
0<\mu_{S}<\frac{2\xi_{C}-1}{-2+\beta+2\xi_{C}},\;\;\;\forall\mu_{S}\in[0,1]
\]

Furthermore, $\frac{\text{d}W}{\text{d}\eta}$ has a nontrivial solution.
That is by studying the sign of the derivative of welfare elections
with respect to $\mu_{S}$ we obtain 
\begin{equation*}
\resizebox{1\textwidth}{!}{
$
\frac{\partial NWF}{\partial\mu_{S}}=-\left[4\beta^{2}\mu_{S}\sigma^{2}+4\beta\left(-1+\mu_{S}\right)\sigma^{2}\left(-1+\xi_{C}\right)+4\beta\mu_{S}\sigma^{2}\left(-1+\xi_{C}\right)+2\left(-1+\mu_{S}\right)\left[\left(1-2\mu\right)^{2}+4\sigma^{2}\right]\left(-1+\xi_{C}\right)^{2}\right]
$
}
\end{equation*}
The sign of the term in brackets is
\[
4\beta^{2}\mu_{S}\sigma^{2}+4\beta\sigma^{2}\left(-1+\xi_{C}\right)\left(-1+2\mu_{S}\right)+2\left(-1+\mu_{S}\right)\left[\left(1-2\mu\right)^{2}+4\sigma^{2}\right]\left(-1+\xi_{C}\right)^{2}>0
\]
\begin{equation*}
\resizebox{1\textwidth}{!}{
$
\mu_{S}\left[4\beta^{2}\sigma^{2}+8\beta\sigma^{2}\left(-1+\xi_{C}\right)+2\left(-1+\xi_{C}\right)^{2}\left[\left(1-2\mu\right)^{2}+4\sigma^{2}\right]\right]>4\beta\sigma^{2}\left(-1+\xi_{C}\right)+2\left(-1+\xi_{C}\right)^{2}\left[\left(1-2\mu\right)^{2}+4\sigma^{2}\right]
$
}
\end{equation*}
\[
\mu_{S}>\underbrace{\frac{4\beta\sigma^{2}\left(-1+\xi_{C}\right)+2\left(-1+\xi_{C}\right)^{2}\left[\left(1-2\mu\right)^{2}+4\sigma^{2}\right]}{4\beta^{2}\sigma^{2}+8\beta\sigma^{2}\left(-1+\xi_{C}\right)+2\left(-1+\xi_{C}\right)^{2}\left[\left(1-2\mu\right)^{2}+4\sigma^{2}\right]}}_{\text{Threshold}<1?}
\]
Saying that the threshold is less than one also means that $\frac{\partial NWF}{\partial\mu_{S}}<0\iff\mu_{S}>\text{Threshold}.$
We want to study this threshold. The conditions given in the text
correspond to such threshold being below $0$ and above, $1$, respectively. 

\[
\begin{cases}
COND1\ensuremath{\rightarrow}\text{Threshold}<0 & \text{Always decreases in time}\\
COND2\ensuremath{\rightarrow}\text{Threshold}>1 & \text{Always increase in time}\\
COND3\ensuremath{\rightarrow}\text{Threshold}\in\left(0,1\right) & \text{Increases first, decreases later}
\end{cases}
\]
COND1 occurs according to the following expression
\[
\beta>\frac{\left(1-\xi_{C}\right)\left(\left(1-2\mu\right)^{2}+4\sigma^{2}\right)}{2\sigma^{2}}
\]

then welfare is always decreasing.\\
For COND2 to occur the numerator of the threshold must be higher than
the denominator. Hence, since there are only two terms differing between
numerator and denominator, the following must be true

\begin{align*}
4\beta\sigma^{2}\left(-1+\xi_{C}\right) & >4\beta^{2}\sigma^{2}+8\beta\sigma^{2}\left(-1+\xi_{C}\right)\\
4\beta\sigma^{2}\left(-1+\xi_{C}\right) & >4\beta\sigma^{2}\left(\beta+2\xi_{C}-2\right)\\
\beta & <1-\xi_{C}
\end{align*}

then welfare is always increasing.\\
Finally, COND3 can be discussed intuitively. Since $\mu_{S}$ is monotonically
decreasing in time, there must be by continuity a $t^{max}$ such
that

\[
\begin{cases}
\frac{\partial W^{N}}{\partial\mu_{S}}<0 & \text{for }t<t^{max}\\
\frac{\partial W^{N}}{\partial\mu_{S}}<0 & \text{for }t>t^{max}
\end{cases}
\]

therefore, $t^{max}$ is a maximum interior of $W^{N}$ when Threshold
$\in\left(0,1\right)$
\end{proof}

\subsection{Proof of Proposition 2}
\begin{proof}
Firstly, define the parameters associated with $\bar{p}$

\[
\bar{p}=\mu_{T}p+\mu_{S}\left(\beta p_{S}+\left(1-\beta\right)p\right)+\mu_{C}\left[1-p+\xi_{C}\left(2p-1\right)\right]
\]

where

\begin{align*}
\alpha_{0} & =\mu_{C}\left(1-\xi_{C}\right)\\
\alpha_{1} & =1-\beta\mu_{S}-2\mu_{C}\left(1-\xi_{C}\right)\\
\alpha_{2} & =\beta\mu_{S}
\end{align*}

Then $\hat{p}$ is given by 

\[
\hat{p}=\frac{\frac{\bar{p}-\left[\alpha_{0}+\alpha_{2}\mu_{y}\right]}{\alpha_{1}}\alpha_{1}^{2}\sigma_{x}^{2}+\alpha_{2}^{2}\sigma_{y}^{2}\mu_{x}}{\alpha_{1}^{2}\sigma_{x}^{2}+\alpha_{2}^{2}\sigma_{y}^{2}}
\]

where

\begin{align*}
\gamma_{0}(t,\lambda) & =\frac{\alpha_{2}^{2}\sigma_{y}^{2}\mu_{x}-\alpha_{2}\alpha_{1}\sigma_{x}^{2}\mu_{y}}{\alpha_{1}^{2}\sigma_{x}^{2}+\alpha_{2}^{2}\sigma_{y}^{2}}\\
\gamma_{1}\left(t,\lambda\right) & =\frac{\alpha_{1}^{2}\sigma_{x}^{2}}{\alpha_{1}^{2}\sigma_{x}^{2}+\alpha_{2}^{2}\sigma_{y}^{2}}\\
\gamma_{2}(t,\lambda) & =\frac{\alpha_{2}\alpha_{1}\sigma_{x}^{2}}{\alpha_{1}^{2}\sigma_{x}^{2}+\alpha_{2}^{2}\sigma_{y}^{2}}
\end{align*}

The bias is zero if and only if the following system has a solution

\[
\begin{cases}
\alpha_{1}=\gamma_{1}\\
\alpha_{2}=\gamma_{2}
\end{cases}
\]

that is 

\[
\begin{cases}
\alpha_{1}=\frac{\alpha_{1}^{2}\sigma_{x}^{2}}{\alpha_{1}^{2}\sigma_{x}^{2}+\alpha_{2}^{2}\sigma_{y}^{2}}\\
\alpha_{2}=\frac{\alpha_{2}\alpha_{1}\sigma_{x}^{2}}{\alpha_{1}^{2}\sigma_{x}^{2}+\alpha_{2}^{2}\sigma_{y}^{2}}
\end{cases}
\]
It is immediate to check that $\gamma_{1}\left(t,\lambda\right)=\alpha_{1}\left(t,\lambda\right)\iff\gamma_{2}\left(t,\lambda\right)=\alpha_{2}\left(t,\lambda\right)$,
so we actually have a single equation and we need to claim that exists
a time such that 

\begin{align*}
\frac{\alpha_{1}^{2}\sigma_{x}^{2}}{\alpha_{1}^{2}\sigma_{x}^{2}+\alpha_{2}^{2}\sigma_{y}^{2}} & =\alpha_{1}\iff\frac{\left(1-\beta\mu_{S}-2\mu_{C}\left(1-\xi_{C}\right)\right)^{2}\sigma_{x}^{2}}{\left(1-\beta\mu_{S}-2\mu_{C}\left(1-\xi_{C}\right)\right)^{2}\sigma_{x}^{2}+\left(\beta\mu_{S}\right)^{2}\sigma_{y}^{2}}=\left(1-\beta\mu_{S}-2\mu_{C}\left(1-\xi_{C}\right)\right)\\
 & \iff\left(1-\beta\mu_{S}-2\mu_{C}\left(1-\xi_{C}\right)\right)\sigma_{x}^{2}=\left(1-\beta\mu_{S}-2\mu_{C}\left(1-\xi_{C}\right)\right)^{2}\sigma_{x}^{2}+\left(\beta\mu_{S}\right)^{2}\sigma_{y}^{2}\\
 & \iff\left(1-\beta\mu_{S}-2\mu_{C}\left(1-\xi_{C}\right)\right)\sigma_{x}^{2}\left(\beta\mu_{S}+2\mu_{C}\left(1-\xi_{C}\right)\right)=\left(\beta\mu_{S}\right)^{2}\sigma_{y}^{2}\\
 & \iff\frac{\left(1-\beta\mu_{S}-2\mu_{C}\left(1-\xi_{C}\right)\right)\left(\beta\mu_{S}+2\mu_{C}\left(1-\xi_{C}\right)\right)}{\left(\beta\mu_{S}\right)^{2}}=\frac{\sigma_{y}^{2}}{\sigma_{x}^{2}}\\
 & \iff\frac{\left(1-\beta\mu_{S}-2\left(1-\mu_{S}\right)\left(1-\xi_{C}\right)\right)\left(\beta\mu_{S}+2\left(1-\mu_{S}\right)\left(1-\xi_{C}\right)\right)}{\left(\beta\mu_{S}\right)^{2}}
\end{align*}

When $\mu_{S}=0$ there cannot be the zero-bias time, because as $t\rightarrow\infty$
this explodes (? can we show this is always increasing in $\mu_{S}$)
because in the limit there is always bias. On the other hand, there
could be a zero-bias time that coincides with $t^{\star}=0$. Indeed,
when $\mu_{S}=1$

\[
\frac{1-\beta}{\beta}=\frac{\sigma_{y}^{2}}{\sigma_{x}^{2}}
\]
An even more special case is when $\xi_{C}=1$. Indeed,

\begin{align*}
\frac{\left(1-\beta\mu_{S}\right)^{2}\sigma_{x}^{2}}{\left(1-\beta\mu_{S}\right)^{2}\sigma_{x}^{2}+\left(\beta\mu_{S}\right)^{2}\sigma_{y}^{2}} & =\left(1-\beta\mu_{S}\right)\\
\left(1-\beta\mu_{S}\right)\sigma_{x}^{2} & =\left(1-\beta\mu_{S}\right)^{2}\sigma_{x}^{2}+\left(\beta\mu_{S}\right)^{2}\sigma_{y}^{2}\\
\left(\frac{1-\beta\mu_{S}}{\beta\mu_{S}}\right) & =\frac{\sigma_{y}^{2}}{\sigma_{x}^{2}}
\end{align*}

Substituting the expression of $\mu_{S}$as a function of $t$ and
$\lambda$

\[
\frac{\sigma_{x}^{2}}{\beta\left(\sigma_{x}^{2}+\sigma_{y}^{2}\right)}=e^{-t\lambda_{1}}
\]

that becomes

\[
t^{\star}=-\frac{1}{\lambda_{1}}\log\left(\frac{\sigma_{x}^{2}}{\beta\left(\sigma_{x}^{2}+\sigma_{y}^{2}\right)}\right)
\]

where the argument of the log must be smaller than 1

\[
\frac{\sigma_{x}^{2}}{\beta\left(\sigma_{x}^{2}+\sigma_{y}^{2}\right)}<1
\]

that is

\[
\frac{\left(1-\beta\right)}{\beta}<\frac{\sigma_{y}^{2}}{\sigma_{x}^{2}}
\]
\end{proof}

\section{Experimental Design Details}

\subsection{Treatments Details and Examples}
In the \textsc{\bfseries newspaper} treatment, the participants are exposed to two news articles: one that is for and one that is against the issue. 
In the \textsc{\bfseries facebook} treatment, participants were exposed to six Facebook posts: two for and two against an issue as well as two irrelevant posts. In treatment \textsc{\bfseries twitter}, participants are exposed to twenty-four tweets: ten for digital privacy, ten against digital privacy, and four irrelevant tweets. Each tweet has an average length of 40 characters, corresponding to 20 words.\footnote{This corresponds to the average length of tweets on twitter.com, see the Appendix for details.}. We give participants 5 seconds to read each tweet before the next one automatically pops until the last one, which is in line with the average reading speed in the US population. In the \textsc{\bfseries partisan twitter} treatment, participants are exposed to 13 tweets: 10 for and 3 irrelevant ones or 10 against and 3 irrelevant ones. Within each treatment, tweets, Facebook posts, and news articles arrive in a random order sequentially (one by screen) and remain on screen for a given fixed amount of time (the participant cannot move to the next screen by him or herself). Each participant is randomly assigned to one of the treatments.

see the online appendix about participant's experimental instructions.

\subsection{Detailed Data Collection}

\paragraph{Preventing duplicates.}Submissions to studies on Prolific are guaranteed to be unique by the firm\footnote{See Prolific unique submission guarantee policy \href{https://researcher-help.prolific.co/hc/en-gb/articles/360009220453-Preventing-participants-from-taking-your-study-multiple-times}{here}.}. Our system is set up such that each participant can have only one submission per study on Prolific. That is, each participant will be listed in your dashboard only once, and can only be paid once. On our side, we also prevent participants to take up several times our experiment in two steps. First, we enable the functionality ``Prevent Ballot Box Stuffing'' which permits to…Second we check participant ID and delete the second submission from the data set of the same ID if we find any. 

\paragraph{Drop-out rates.} Here put the drop out (or in the main text). 

\paragraph{High vs low-quality submissions.}Participants joining the Prolific pool receive a rate based on the quality of their engagement with the studies. If they are rejected from a study then they receive a malus. If they receive too much malus, then they are removed by the pool from the company\footnote{See Prolific pool removal Policy \href{https://researcher-help.prolific.co/hc/en-gb/articles/360009092394-Reviewing-submissions-How-do-I-decide-who-to-accept-reject-}{here}.}. Based on this long term contract, participants are incentivized to pay attention and follow the expectations of each study. Hence, a good research behavior has emerged on Prolific according to which, participants themselves can vol voluntarily withdraw their submissions if they feel they did a mistake such as rushing too much, letting the survey opened for a long period of time without engaging with it, and so on\footnote{See Prolific update regarding this behavior \href{https://researcher-help.prolific.co/hc/en-gb/articles/360009092394-Reviewing-submissions-How-do-I-decide-who-to-accept-reject-}{here}.}. According to these standards, we kept submissions rejections as low as possible, following standard in online experimental economics. Participants who fail at least one fair attention check are rejected and not paid. Following Prolific standards, participants who are statistical outliers (3 standard deviations below the mean) are excluded from the good complete data set. 

\paragraph{Payments and communication.}We make sure to review participants’ submissions within within 24-48 hours after they have completed the study. This means that within this time frame, if we accept their submission, they receive their fixed and bonus payment. Otherwise, we reject their submissions and send to them a personalized e-mail(\footnote{Partially-anonymized through Prolific messaging app which put the researcher’s name visible to the participants and only the participants ID visible to the researcher.}), detailing the reason of the rejection, leaving participants the opportunity to contact us afterwards if they firmly believe the decision to be unfair (motivate their perspective). Participants can also contact us at any time if they encounter problems with our study or just have questions about it. 

\subsection{Detailed Elicitations}

\subsubsection{Political Preferences}

\paragraph{Ex-ante and ex-post political preferences.} Before the  treatment, we survey participants' preferences (self-reported) on political issues: guns, crime, climate, welfare, and digital privacy issues. We use the standard congressional metrics, including digital issues. We elicit more than only digital preferences to ensure that participants do not guess at this stage which preferences we focus on in the remaining of the experiment (treatment and critical thinking essay), to minimize their social desirability bias. After the  treatment on digital privacy, we survey again participants to elicit their preferences about digital privacy. We use the following scale. 

\begin{enumerate}
    \item On the issue of gun regulation, do you support or oppose the following proposal?
    \item On the issue of environmental policies, do you support or oppose the following proposal?
    \item On the issue of crime policies, do you support or oppose the following proposal?
    \item On the issue of digital policies, do you support or oppose each of the following proposals?
\end{enumerate}

\subsubsection{Digital Knowledge Test}

see the online appendix about participant's experimental instructions.

\subsubsection{Issue Familiarity}
\begin{enumerate}
    \item In the remainder of the experiment, we will focus on the following political issue. Please state again your preference.
    \item Have you thought deeply about this issue before participating in this study? [Yes/No]
\end{enumerate}

\subsubsection{Listing Reasons}
If $y$es to the previous question, then participants see this question: 

You answered "Yes" to the previous question. you will be asked now to provide, at most, two reasons which justify your position and two reasons which justify the opposite position. If you do not know any reasons, please select "I am unable to list any logical reason at the moment". you do not need to agree with these reasons: they just need to be a logical justification for or against your position. your payment WILL NOT depend on your answer to this question. However, your honest answer is of paramount importance for the success of this study. 

\begin{enumerate}
    \item Reasons which justify your position
    \begin{itemize}
        \item Reason 1: [write text here]
        \item Reason 2: [write text here]
        \item I am unable to list any logical reason at the moment  
    \end{itemize}
       \item Reasons which oppose your position
    \begin{itemize}
        \item Reason 1: [write text here]
        \item Reason 2: [write text here]
        \item I am unable to list any logical reason at the moment 
    \end{itemize}
    
\end{enumerate}

\subsubsection{Internal Uncertainty}
How certain are you of your preference regarding the digital privacy issue? By "Certain", we mean that you feel confident enough to vote for your political preference if asked to you in a real life political committee. Select among the following options:  
\begin{itemize}
    \item Completely Uncertain  
    \item Rather Uncertainty
    \item Rather Certain
    \item Completely Certain
\end{itemize}

\subsubsection{Need for Cognition}
For each sentence below, please select how uncharacteristic or characteristic this is for you personally.

\subsubsection{Cognitive Flexibility}

\subsubsection{Habits of News Consumptopn}

see the online appendix about participant's experimental instructions.

\subsection{Detailed Description of Graders' Instructions}

We recruited 20 psychologists (doctoral level or above) specializing in cognitive psychology at Princeton University. Each grader was randomly assigned a ``grading treatment'' (i.e., a set of essays to grade). Such a set of essays was randomly built, containing essays from all four  treatments. Additionally, graders were not informed about which  treatment the subjects were assigned. Psychologists must grade a very short paragraph (around 300 words or fewer) as follows. The grading consists of giving a passing grade if psychologists judge that the participant ``realizes that the issue is ambivalent,'' a failing grade otherwise. What may happen is to confound high cognitive sophistication (i.e., the ability to write well-written essays in English), facilitated by the fact that they read some arguments right before this essay exercise with their self-reasoning skill "realizing that the issue is ambivalent", which is the variable that we want to elicit. This is a specific case that is still challenging for AI-based grading software and the main reason why human expertise is uniquely useful.

We define ``realizing that the issue is ambivalent'' as the awareness of an individual to recognize that there can be perfectly logical but opposite arguments in favor of and against the same issue that renders the decision-making process complex. Such attitudinal ambivalence leads to temporarily conflicting preferences; namely, one preference for the issue at hand and one preference against the issue at hand. There are different ways of measuring this ``awareness,'' as documented in the social psychology and cognitive psychology literature. In our study, we capture this awareness by observing individuals reasoning and elaborating in a personal way on the pros and cons of the same issue in a textual format. 

Each grader was paid a fixed fee of \$50 for each grading session. Each grader could participate up to three times in our experiment, and no grader could be assigned twice to the same grading treatment. For robustness, each essay was corrected three times by different psychologists. Despite ``triple-eliciting'' such grades, this metric can still be prone to measurement error. Accordingly, we suggest interpreting the estimated \emph{levels} of $\lambda$ with caution. However, our focus is on the difference between the  treatments. Therefore, such measurement error does not affect this difference. 

see the online appendix about participant's experimental instructions.

\subsection{Heterogeneous Critical Thinking Classification}

\subsubsection{Critical Thinking Classification Results}

Table \ref{tab:3classresults} shows the classification results of of individuals as \textit{S}tereotype and \textit{A}ware.

\begin{table}[H]
\centering
\footnotesize
\addtolength{\tabcolsep}{5pt} 
\begin{tabular}{lccc}
\toprule \toprule
Treatment &\multicolumn{1}{c}{$S_{0}\rightarrow S_{1}$}& \multicolumn{1}{c}{$S_{0} \rightarrow A_{1}$}  & \multicolumn{1}{c}{$A_{0}\rightarrow A_{1}$}  \\
\midrule
\sc{newspaper} &  111 &     49 &  12      \\
\sc{twitter}   &    135&   43  & 15         \\
\sc{facebook}   &     111 &    60    &   11           \\
\midrule
$N$ & 357 &   152 & 38 \\
\bottomrule \bottomrule
\end{tabular}
\caption{\sc Table 2: Classification \textbf{results} before/after treatment}
\label{tab:3classresults}
\end{table}

\subsubsection{Awareness With Cognitive Styles Heterogeneity with Unanimity}

Awareness With Cognitive Flexibility, with Unanimity

        \begin{table}[H]
        \begin{tabular}{lccc}
        \toprule \toprule
        Treatment &\multicolumn{1}{c}{\sc{NP}} & \multicolumn{1}{c}{\sc{twitter}}  & \multicolumn{1}{c}{\sc{facebook}}  \\
        \midrule
        \sc{newspaper} & $\cdot$ & 1.301 & 1.661     \\
        &  & (0.091) & (0.095) \\
        \sc{twitter}   &  $\cdot$ & $\cdot$ & 0.422         \\
        &  &  & (0.093) \\
        \sc{facebook}   & $\cdot$ & $\cdot$ & $\cdot$           \\
        \bottomrule \bottomrule
        \multicolumn{3}{l}{\footnotesize Standard errors in parentheses}\\
        \multicolumn{3}{l}{\footnotesize \sym{*} \(p<0.05\), \sym{**} \(p<0.01\), \sym{***} \(p<0.001\)}\\
        \end{tabular}
        \caption{\sc \textit{t-ratio} for \textit{High Need for Cognition} with Unanimity}
        \end{table}\label{tab:xx}


         \begin{table}[H]
        \begin{tabular}{lccc}
        \toprule \toprule
        Treatment &\multicolumn{1}{c}{\sc{NP}} & \multicolumn{1}{c}{\sc{twitter}}  & \multicolumn{1}{c}{\sc{facebook}}  \\
        \midrule
        \sc{newspaper} & $\cdot$ & 0.320 & -0.388     \\
        &  & (0.066) & (0.065) \\
        \sc{twitter}   &  $\cdot$ & $\cdot$ & -0.965         \\
        &  &  & (0.064) \\
        \sc{facebook}   & $\cdot$ & $\cdot$ & $\cdot$           \\
        \bottomrule \bottomrule
        \multicolumn{3}{l}{\footnotesize Standard errors in parentheses}\\
        \multicolumn{3}{l}{\footnotesize \sym{*} \(p<0.05\), \sym{**} \(p<0.01\), \sym{***} \(p<0.001\)}\\
           \end{tabular}
            \caption{\sc Table 6: \textit{t-ratio} for \textit{Low Need for Cognition} with Unanimity}
     \end{table}\label{tab:xxx}

Awareness With Need for Cognition, with Unanimity

\begin{table}[H]

        \begin{tabular}{lccc}
        \toprule \toprule
        Treatment &\multicolumn{1}{c}{\sc{NP}} & \multicolumn{1}{c}{\sc{twitter}}  & \multicolumn{1}{c}{\sc{facebook}}  \\
        \midrule
        \sc{newspaper} & $\cdot$ & 1.061 & -2.238*     \\
        &  & (0.084) & (0.084) \\
        \sc{twitter}   &  $\cdot$ & $\cdot$ & -1.300         \\
        &  &  & (0.083) \\
        \sc{facebook}   & $\cdot$ & $\cdot$ & $\cdot$           \\
        \bottomrule \bottomrule
        \multicolumn{3}{l}{\footnotesize Standard errors in parentheses}\\
        \multicolumn{3}{l}{\footnotesize \sym{*} \(p<0.05\), \sym{**} \(p<0.01\), \sym{***} \(p<0.001\)}\\
        \end{tabular}
        \caption{\sc Table 6: \textit{t-ratio} for \textit{High Need for Cognition} with Unanimity} 
\end{table}
\begin{table}[H]
        \begin{tabular}{lccc}
        \toprule \toprule
        Treatment &\multicolumn{1}{c}{\sc{NP}} & \multicolumn{1}{c}{\sc{twitter}}  & \multicolumn{1}{c}{\sc{facebook}}  \\
        \midrule
        \sc{newspaper} & $\cdot$ & 0.455 & 0.705     \\
        &  & (0.069) & (0.069) \\
        \sc{twitter}   &  $\cdot$ & $\cdot$ & 0.258         \\
        &  &  & (0.069) \\
        \sc{facebook}   & $\cdot$ & $\cdot$ & $\cdot$           \\
        \bottomrule \bottomrule
        \multicolumn{3}{l}{\footnotesize Standard errors in parentheses}\\
        \multicolumn{3}{l}{\sym{*} \(p<0.05\), \sym{**} \(p<0.01\), \sym{***} \(p<0.001\)}\\
        \end{tabular}
        \caption{\sc Table 6: \textit{t-ratio} for \textit{Low Need for Cognition} with Unanimity}
\end{table}

\subsubsection{Awareness With Cognitive Styles Heterogeneity with Majority}

Awareness With Cognitive Flexibility, with Majority
\begin{table}[H]
        \begin{tabular}{lccc}
        \toprule \toprule
        Treatment &\multicolumn{1}{c}{\sc{NP}} & \multicolumn{1}{c}{\sc{twitter}}  & \multicolumn{1}{c}{\sc{facebook}}  \\
        \midrule
        \sc{newspaper} & $\cdot$ & 0.658 & -0.924      \\
        &  & (0.076) & (0.087) \\
        \sc{twitter}   &  $\cdot$ & $\cdot$ & -1.609         \\
        &  &  & (0.081) \\
        \sc{facebook}   & $\cdot$ & $\cdot$ & $\cdot$           \\
        \bottomrule \bottomrule
        \multicolumn{3}{l}{\footnotesize Standard errors in parentheses}\\
        \multicolumn{3}{l}{\footnotesize \sym{*} \(p<0.05\), \sym{**} \(p<0.01\), \sym{***} \(p<0.001\)}\\
        \end{tabular}
        \caption{\sc Table 6: \textit{t-ratio} for \textit{High Flexibility}}
  \end{table}
\begin{table}[H]        
    \begin{tabular}{lccc}
    \toprule \toprule
    Treatment &\multicolumn{1}{c}{\sc{NP}} & \multicolumn{1}{c}{\sc{twitter}}  & \multicolumn{1}{c}{\sc{facebook}}  \\
    \midrule
    \sc{newspaper} & $\cdot$ & 1.132 & -0.388      \\
    &  & (0.062) & (0.064) \\
    \sc{twitter}   &  $\cdot$ & $\cdot$ & -1.564         \\
    &  &  & (0.061) \\
    \sc{facebook}   & $\cdot$ & $\cdot$ & $\cdot$           \\
    \bottomrule \bottomrule
    \multicolumn{3}{l}{\footnotesize Standard errors in parentheses}\\
    \multicolumn{3}{l}{\footnotesize \sym{*} \(p<0.05\), \sym{**} \(p<0.01\), \sym{***} \(p<0.001\)}\\
    \end{tabular}
    \caption{\sc Table 7: \textit{t-ratio} for \textit{Low Flexibility}}
\end{table}

Awareness With Need for Cognition, with Majority

\begin{table}[H] 
        \begin{tabular}{lccc}
        \toprule \toprule
        Treatment &\multicolumn{1}{c}{\sc{newspaper}} & \multicolumn{1}{c}{\sc{twitter}}  & \multicolumn{1}{c}{\sc{facebook}}  \\
        \midrule
        \sc{newspaper} & $\cdot$ & 0.764 & -2.238*      \\
        &  & (0.070) & (0.079) \\
        \sc{twitter}   &  $\cdot$ & $\cdot$ & -3.087**         \\
        &  &  & (0.075) \\
        \sc{facebook}   & $\cdot$ & $\cdot$ & $\cdot$           \\
        \bottomrule \bottomrule
        \multicolumn{3}{l}{\footnotesize Standard errors in parentheses}\\
        \multicolumn{3}{l}{\footnotesize \sym{*} \(p<0.05\), \sym{**} \(p<0.01\), \sym{***} \(p<0.001\)}\\
        \end{tabular}
        \caption{\sc Table 4: \textit{t-ratio} for \textbf{high} need for cognition}
        \label{tab:highneed}
  \end{table}  

    \begin{table}[H]
    \begin{tabular}{lccc}
    \toprule \toprule
    Treatment &\multicolumn{1}{c}{\sc{newspaper}} & \multicolumn{1}{c}{\sc{twitter}}  & \multicolumn{1}{c}{\sc{facebook}}  \\
    \midrule
    \sc{newspaper} & $\cdot$ & 1.094 & 0.703      \\
    &  & (0.066) & (0.067) \\
    \sc{twitter}   &  $\cdot$ & $\cdot$ & -0.396         \\
    &  &  & (0.063) \\
    \sc{facebook}   & $\cdot$ & $\cdot$ & $\cdot$           \\
    \bottomrule \bottomrule
    \multicolumn{3}{l}{\footnotesize Standard errors in parentheses}\\
    \multicolumn{3}{l}{\footnotesize \sym{*} \(p<0.05\), \sym{**} \(p<0.01\), \sym{***} \(p<0.001\)}\\
    \end{tabular}
        \caption{\sc Table 5: \textit{t-ratio} for low need for cognition}
    \label{tab:lowneed}
\end{table}

\subsection{Threshold changes}

 \begin{table}[H]
    \centering
    \footnotesize
    \addtolength{\tabcolsep}{5pt} 
    \begin{tabular}{lccc}
    \toprule \toprule
    Treatment &\multicolumn{1}{c}{\sc{newspaper}} & \multicolumn{1}{c}{\sc{twitter}}  & \multicolumn{1}{c}{\sc{facebook}}  \\
    \midrule
    \sc{newspaper} & $\cdot$ & 1.278 & -0.923      \\
    &  & (0.054) & (0.054) \\
    {\sc{twitter}}    &  $\cdot$ & $\cdot$ & -2.262\sym{*}       \\
    &  &  & (0.053) \\
    \sc{facebook} & $\cdot$ & $\cdot$ & $\cdot$ \\ \bottomrule \bottomrule
    \multicolumn{3}{l}{\footnotesize Standard errors in parentheses}\\
    \multicolumn{3}{l}{\footnotesize \sym{*} \(p<0.05\), \sym{**} \(p<0.01\), \sym{***} \(p<0.001\)}\\
    \end{tabular}
    \caption{\sc \textit{t-ratio} difference-in-means with threshold of KTS = 8 and Reason Counter = 2}
    \label{tab:threshold82}
\end{table}

 \begin{table}[H]
    \centering
    \footnotesize
    \addtolength{\tabcolsep}{5pt} 
    \begin{tabular}{lccc}
    \toprule \toprule
    Treatment &\multicolumn{1}{c}{\sc{newspaper}} & \multicolumn{1}{c}{\sc{twitter}}  & \multicolumn{1}{c}{\sc{facebook}}  \\
    \midrule
    \sc{newspaper} & $\cdot$ & 1.222 & -0.799      \\
    &  & (0.054) & (0.054) \\
    {\sc{twitter}}    &  $\cdot$ & $\cdot$ & -2.067*        \\
    &  &  & (0.053) \\
    \sc{facebook} & $\cdot$ & $\cdot$ & $\cdot$ \\ \bottomrule \bottomrule
    \multicolumn{3}{l}{\footnotesize Standard errors in parentheses}\\
    \multicolumn{3}{l}{\footnotesize \sym{*} \(p<0.05\), \sym{**} \(p<0.01\), \sym{***} \(p<0.001\)}\\
    \end{tabular}
    \caption{\sc \textit{t-ratio} difference-in-means with threshold of KTS = 7 and Reason Counter = 3}
    \label{tab:threshold73}
\end{table}

 \begin{table}[H]
    \centering
    \footnotesize
    \addtolength{\tabcolsep}{5pt} 
    \begin{tabular}{lccc}
    \toprule \toprule
    Treatment &\multicolumn{1}{c}{\sc{newspaper}} & \multicolumn{1}{c}{\sc{twitter}}  & \multicolumn{1}{c}{\sc{facebook}}  \\
    \midrule
    \sc{newspaper} & $\cdot$ & 1.136 & -0.703      \\
    &  & (0.049) & (0.052) \\
    {\sc{twitter}}    &  $\cdot$ & $\cdot$ & -1.985*        \\
    &  &  & (0.049) \\
    \sc{facebook} & $\cdot$ & $\cdot$ & $\cdot$ \\ \bottomrule \bottomrule
    \multicolumn{3}{l}{\footnotesize Standard errors in parentheses}\\
    \multicolumn{3}{l}{\footnotesize \sym{*} \(p<0.05\), \sym{**} \(p<0.01\), \sym{***} \(p<0.001\)}\\
    \end{tabular}
    \caption{\sc \textit{t-ratio} difference-in-means with threshold of KTS = 7 and Reason Counter = 2}
    \label{tab:threshold72}
\end{table}

 \begin{table}[H]
    \centering
    \footnotesize
    \addtolength{\tabcolsep}{5pt} 
    \begin{tabular}{lccc}
    \toprule \toprule
    Treatment &\multicolumn{1}{c}{\sc{newspaper}} & \multicolumn{1}{c}{\sc{twitter}}  & \multicolumn{1}{c}{\sc{facebook}}  \\
    \midrule
    \sc{newspaper} & $\cdot$ & 1.428 & -0.507      \\
    &  & (0.049) & (0.052) \\
    {\sc{twitter}}    &  $\cdot$ & $\cdot$ & -1.985*        \\
    &  &  & (0.049) \\
    \sc{facebook} & $\cdot$ & $\cdot$ & $\cdot$ \\ \bottomrule \bottomrule
    \multicolumn{3}{l}{\footnotesize Standard errors in parentheses}\\
    \multicolumn{3}{l}{\footnotesize \sym{*} \(p<0.05\), \sym{**} \(p<0.01\), \sym{***} \(p<0.001\)}\\
    \end{tabular}
    \caption{\sc \textit{t-ratio} difference-in-means with threshold of KTS = 6 and Reason Counter = 3}
    \label{tab:threshold63}
\end{table}

\subsection{AI Digital Grade}

The last robustness check that we perform is to split the sample conditioning on the the grade of the essay on the digital topic evaluated by the algorithm of the AI.Indeed, the essay is written after undertaking the experiment and it might influence the writing quality of the essay. In particular, our reasoning is that if no difference in proportion is statistically significant, this means that there is no systematic difference between those who were treated through Facebook and those through Twitter, and indeed from Table 17 this is the case.

 \begin{table}[H]
    \centering
    \footnotesize
    \addtolength{\tabcolsep}{5pt} 
    \begin{tabular}{lccc}
    \toprule \toprule
    Treatment &\multicolumn{1}{c}{\sc{newspaper}} & \multicolumn{1}{c}{\sc{twitter}}  & \multicolumn{1}{c}{\sc{facebook}}  \\
    \midrule
    \sc{newspaper} & $\cdot$ & 1.716 & 0.234      \\
    &  & (0.078) & (0.081) \\
    {\sc{twitter}}    &  $\cdot$ & $\cdot$ & -1.565        \\
    &  &  & (0.074) \\
    \sc{facebook} & $\cdot$ & $\cdot$ & $\cdot$ \\ \bottomrule \bottomrule
    \multicolumn{3}{l}{\footnotesize Standard errors in parentheses}\\
    \multicolumn{3}{l}{\footnotesize \sym{*} \(p<0.05\), \sym{**} \(p<0.01\), \sym{***} \(p<0.001\)}\\
    \end{tabular}
    \caption{\sc \textit{t-ratio} with High AI Digital Grades}
    \label{tab:grammarlyhigh}
\end{table}

 \begin{table}[H]
    \centering
    \footnotesize
    \addtolength{\tabcolsep}{5pt} 
    \begin{tabular}{lccc}
    \toprule \toprule
    Treatment &\multicolumn{1}{c}{\sc{newspaper}} & \multicolumn{1}{c}{\sc{twitter}}  & \multicolumn{1}{c}{\sc{facebook}}  \\
    \midrule
    \sc{newspaper} & $\cdot$ & 0.098 & -1.093      \\
    &  & (0.061) & (0.067) \\
    {\sc{twitter}}    &  $\cdot$ & $\cdot$ & -1.207        \\
    &  &  & (0.065) \\
    \sc{facebook} & $\cdot$ & $\cdot$ & $\cdot$ \\ \bottomrule \bottomrule
    \multicolumn{3}{l}{\footnotesize Standard errors in parentheses}\\
    \multicolumn{3}{l}{\footnotesize \sym{*} \(p<0.05\), \sym{**} \(p<0.01\), \sym{***} \(p<0.001\)}\\
    \end{tabular}
    \caption{\sc \textit{t-ratio} with Low AI Digital Grades}
    \label{tab:grammarlylow}
\end{table}

\section{Model With Three-State Critical Thinking}

We propose an additional model where agents can be at three different states of critical thinking: not engaging with critical thinking, performing critical thinking (either in its first or second state), and having finished performing critical thinking. In our two-stage model, we considered performing critical thinking as having finished performing it. In this scenario, we propose a three-stage (not fully identified) model that considers the three stages distinctively.

In this economy, the object of interest is the distribution of stable preferences over a binary policy space in a large population. Each individual $j$ inside the population is characterized by a three-dimensional type
\[
\left(x_{j},y_{j},i_{j}\right)\in\mathcal{J}\coloneqq\left\{ 0,1\right\} \times\left\{ 0,1\right\} \times\left\{0,1\right\}
\]
where $x_j$, represents the stereotypical preference individual $j$ would self-report when presented with an ambivalent issue for the first time -- that is, by definition, before undergoing a critical thinking phase; $y_j$ differs potentially from $x_j$ as it represents the stable preference that $j$ holds after completing their period of critical thinking; the cognitive type $i_j$ refers to the cognitive type $i_j$, interacting with the  format, determines how easily individual $j$ moves into (and out of) critical thinking. 

Individuals go through a three-step process of "critical thinking" as they form their preferences. The process begins with a "stereotypical-self" state, followed by a period of critical thinking, and ultimately leading to a "stable-self" state. We assume that this process is irreversible and that once individuals reach a stable-self state, they no longer question their preferences. There is no additional "information" that has to come and change their worldview: the process of critical thinking provides a final and stable answer to ambivalent issues. When asked to report their preferences on a policy issue, individuals in either their stereotypical self or stable self state will vote according to their respective preferences, $x_j,y_j$, respectively. Those who are still in the critical thinking phase will abstain from voting. 

The transition between the different phases is determined by an individual's cognitive style and the characteristics of the storytelling format. Hence, the  storytelling format is instrumental in the agent's transition from a stereotypical state to the stable one. By constructing our model, this transition is captured by the critical thinking phase. An economy of stable preferences is preferable from efficiency and welfare perspectives to an economy of stereotypical preferences. We formally present such an economy below.

\subsection{Model Identification}

Using reported preferences of individuals that do the $S (Stereotype) \to T (Type) $ transition (i.e. we observe ex ante $x_{S}$ then $y$), we get 
\[
\mathbb{E}\left[x_{S}\left|y=1\right.\right]=\left(1-\beta\right)+\beta p_{S}
\]
\[
\mathbb{E}\left[x_{S}\left|y=0\right.\right]=\beta p_{S}
\]
which gives the estimators
\[
\hat{\beta}=1-\left(\bar{x}_{S\left|1\right.}-\bar{x}_{S\left|0\right.}\right)
\]
and 
\[
\hat{p}_{S}=\frac{\bar{x}_{S\left|0\right.}}{\hat{\beta}}
\]
clearly $\hat{p}=\bar{y}$. Finally, using the reported preferences of
individuals that do the $A \to T (Type)$ transition we can estimate $\xi_{A} $
as 
\[
\mathbb{E}\left[x_{A}\left|y=1\right.\right]=\xi_{A}
\]
\[
\mathbb{E}\left[x_{A}\left|y=0\right.\right]=1-\xi_{A}
\]
so $\hat{\xi}_{A}=\bar{x}_{NU\left|1\right.}$ or $\hat{\hat{\xi}}_{A}=1-\bar{x}_{NU\left|0\right.}$.
Notice that we can test the assumed symmetry by testing that $\hat{\xi}_{A}=\hat{\hat{\xi}}_{A}$.
Since in our dataset we have few agents that start in $A$ this test has almost no power. 

\subsection{General Results}

The basic decomposition 
\begin{align*}
W_{E} & =W_{P}+Bias\\
-\mathbb{E}\left[\left(p-\bar{p}\right)^{2}\right] & =-\left(\mathbb{E}\left[\left(p-\hat{p}\right)^{2}\right]+\mathbb{E}\left[\left(\hat{p}-\bar{p}\right)^{2}\right]\right)
\end{align*}
is still clearly valid. However, $\bar{p}$ is now given by

\begin{align*}
\bar{p} & =\mu_{T}p+\mu_{S}\left(\beta p_{S}+\left(1-\beta\right)p\right)+\mu_{A}\left[1-p+\xi_{A}\left(2p-1\right)\right]\\
 & =\alpha_{0}\left(t,\lambda\right)+\alpha_{1}\left(t,\lambda\right)p+\alpha_{2}\left(t,\lambda\right)p_{S}
\end{align*}
with

\begin{align*}
\alpha_{0} & =\mu_{A}\left(1-\xi_{A}\right)\\
\alpha_{1} & =1-\beta\mu_{S}-2\mu_{A}\left(1-\xi_{A}\right)\\
\alpha_{2} & =\beta\mu_{S}
\end{align*}
where $\hat{p}$ (that was wrong in the previous file since for non-normal
random variables we do not know the expectation of $p$ given the
convex combination $\beta p_{s}+\left(1-\beta\right)p$) is given
by 
\[
\hat{p}=\frac{\frac{\bar{p}-\left[\alpha_{0}+\alpha_{2}\mu_{y}\right]}{\alpha_{1}}\alpha_{1}^{2}\sigma_{x}^{2}+\alpha_{2}^{2}\sigma_{y}^{2}\mu_{x}}{\alpha_{1}^{2}\sigma_{x}^{2}+\alpha_{2}^{2}\sigma_{y}^{2}}=\tilde{\alpha_{0}}\left(t,\lambda\right)+\tilde{\alpha_{1}}\left(t,\lambda\right)p+\tilde{\alpha_{2}}\left(t,\lambda\right)p_{S}
\]
\[
\frac{\frac{\alpha_{0}\left(t,\lambda\right)+\alpha_{1}\left(t,\lambda\right)p+\alpha_{2}\left(t,\lambda\right)p_{S}-\left[\alpha_{0}+\alpha_{2}\mu_{y}\right]}{\alpha_{1}}\alpha_{1}^{2}\sigma_{x}^{2}+\alpha_{2}^{2}\sigma_{y}^{2}\mu_{x}}{\alpha_{1}^{2}\sigma_{x}^{2}+\alpha_{2}^{2}\sigma_{y}^{2}}
\]
 so 
\[
\tilde{\alpha_{1}}\left(t,\lambda\right)=\frac{\alpha_{1}^{2}\sigma_{x}^{2}}{\alpha_{1}^{2}\sigma_{x}^{2}+\alpha_{2}^{2}\sigma_{y}^{2}}
\]
\[
\tilde{\alpha_{2}}\left(t,\lambda\right)=\frac{\alpha_{2}\alpha_{1}\sigma_{x}^{2}}{\alpha_{1}^{2}\sigma_{x}^{2}+\alpha_{2}^{2}\sigma_{y}^{2}}
\]
when is it 
\begin{align*}
\tilde{\alpha_{1}}\left(t,\lambda\right) & =\alpha_{1}\left(t,\lambda\right)\iff\frac{\alpha_{1}^{2}\sigma_{x}^{2}}{\alpha_{1}^{2}\sigma_{x}^{2}+\alpha_{2}^{2}\sigma_{y}^{2}}=\alpha_{1}\iff\\
\text{Same \ensuremath{\sigma}} & =\alpha_{1}\left(1-\alpha_{1}\right)=\alpha_{2}^{2}\iff\left(1-\beta\mu_{S}\right)\left(\beta\mu_{S}\right)=\left(\beta\mu_{S}\right)^{2}
\end{align*}
It is immediate to check that $\tilde{\alpha_{1}}\left(t,\lambda\right)=\alpha_{1}\left(t,\lambda\right)\iff\tilde{\alpha_{2}}\left(t,\lambda\right)=\alpha_{2}\left(t,\lambda\right)$
so we actually have a single equation and we need to claim that $\exists$
time such that 
\begin{align*}
\frac{\alpha_{1}^{2}\sigma_{x}^{2}}{\alpha_{1}^{2}\sigma_{x}^{2}+\alpha_{2}^{2}\sigma_{y}^{2}} & =\alpha_{1}\iff\frac{\left(1-\beta\mu_{S}-2\mu_{A}\left(1-\xi_{A}\right)\right)^{2}\sigma_{x}^{2}}{\left(1-\beta\mu_{S}-2\mu_{A}\left(1-\xi_{A}\right)\right)^{2}\sigma_{x}^{2}+\left(\beta\mu_{S}\right)^{2}\sigma_{y}^{2}}=\left(1-\beta\mu_{S}-2\mu_{A}\left(1-\xi_{A}\right)\right)\\
 & \iff\left(1-\beta\mu_{S}-2\mu_{A}\left(1-\xi_{A}\right)\right)\sigma_{x}^{2}=\left(1-\beta\mu_{S}-2\mu_{A}\left(1-\xi_{A}\right)\right)^{2}\sigma_{x}^{2}+\left(\beta\mu_{S}\right)^{2}\sigma_{y}^{2}\\
 & \iff\left(1-\beta\mu_{S}-2\mu_{A}\left(1-\xi_{A}\right)\right)\sigma_{x}^{2}\left(\beta\mu_{S}+2\mu_{A}\left(1-\xi_{A}\right)\right)=\left(\beta\mu_{S}\right)^{2}\sigma_{y}^{2}\\
 & \iff\frac{\left(1-\beta\mu_{S}-2\mu_{A}\left(1-\xi_{A}\right)\right)\left(\beta\mu_{S}+2\mu_{A}\left(1-\xi_{A}\right)\right)}{\left(\beta\mu_{S}\right)^{2}}=\frac{\sigma_{y}^{2}}{\sigma_{x}^{2}}\\
\end{align*}
now substituting $\mu_{S},\mu_{A}$ we have the LHS is increasing
to $\infty$ in $t$, therefore there is a unique solution provided
that it starts below $\frac{\sigma_{y}^{2}}{\sigma_{x}^{2}}$, that
is if $\frac{1-\beta}{\beta}<\frac{\sigma_{y}^{2}}{\sigma_{x}^{2}}$
($\beta$ is large enough)

as $\mu_{S}\to0$, this explodes (there is always bias in the
limit), while at the beginning there is zero bias iff 
\[
\frac{1-\beta}{\beta}=\frac{\sigma_{y}^{2}}{\sigma_{x}^{2}}
\]

{[}example, $\sigma_{x}^{2}=\frac{1}{2},\sigma_{y}^{2}=\frac{1}{6}$
$\beta=\frac{3}{4}\implies\frac{1-\beta}{\beta}=\frac{1}{3}$ {]}

this is the zero-bias time. Even more special cases $\xi_{A}=1$
\[
\frac{\left(1-\beta\mu_{S}\right)^{2}\sigma_{x}^{2}}{\left(1-\beta\mu_{S}\right)^{2}\sigma_{x}^{2}+\left(\beta\mu_{S}\right)^{2}\sigma_{y}^{2}}=\left(1-\beta\mu_{S}\right)
\]
\[
\left(1-\beta\mu_{S}\right)\sigma_{x}^{2}=\left(1-\beta\mu_{S}\right)^{2}\sigma_{x}^{2}+\left(\beta\mu_{S}\right)^{2}\sigma_{y}^{2}
\]
\[
\left(\frac{1-\beta\mu_{S}}{\beta\mu_{S}}\right)=\frac{\sigma_{y}^{2}}{\sigma_{x}^{2}}
\]
Since the LHS is decreasing in $\mu_{S}$ and the RHS is increasing,
then there is at most one solution. It has none if 
\[
\frac{1-\beta}{\beta}>\frac{\sigma_{y}^{2}}{\sigma_{x}^{2}}
\]

Furthermore we get 
\begin{align*}
W_{P} & =-\mathbb{E}\left[\left(p-\hat{p}\right)^{2}\right]=-\frac{\alpha_{2}^{2}\sigma_{y}^{2}\sigma_{x}^{2}}{\alpha_{1}^{2}\sigma_{x}^{2}+\alpha_{2}^{2}\sigma_{y}^{2}}\\
\text{If }\sigma_{x}^{2}=\sigma_{y}^{2}=\sigma^{2} & =-\frac{\alpha_{2}^{2}}{\alpha_{1}^{2}+\alpha_{2}^{2}}\sigma^{2}\\
\xi_{A}=1 & =\frac{\left(\beta\mu_{S}\right)^{2}}{2\beta\mu_{S}\left[1-\beta\mu_{S}\right]+1}\sigma^{2}??
\end{align*}

\subsubsection*{Aside: No Bias}

Condition for no bias is that coefficients in $\bar{p}$ are the same
as in $\hat{p}$ that is, 
\[
If \exists t:B\left(t\right)=0
\]
\[
\alpha_{0}+\alpha_{1}p+\alpha_{2}p_{S}
\]
\[
\frac{\frac{\alpha_{1}p+\alpha_{2}p_{S}-\alpha_{2}\mu_{y}}{\alpha_{1}}\alpha_{1}^{2}\sigma_{x}^{2}+\alpha_{2}^{2}\sigma_{y}^{2}\mu_{x}}{\alpha_{1}^{2}\sigma_{x}^{2}+\alpha_{2}^{2}\sigma_{y}^{2}}=\frac{\alpha_{2}^{2}\sigma_{y}^{2}\mu_{x}-\alpha_{2}\alpha_{1}\sigma_{x}^{2}\mu_{y}}{\alpha_{1}^{2}\sigma_{x}^{2}+\alpha_{2}^{2}\sigma_{y}^{2}}+\frac{\alpha_{1}^{2}\sigma_{x}^{2}}{\alpha_{1}^{2}\sigma_{x}^{2}+\alpha_{2}^{2}\sigma_{y}^{2}}p+\frac{\alpha_{2}\alpha_{1}\sigma_{x}^{2}}{\alpha_{1}^{2}\sigma_{x}^{2}+\alpha_{2}^{2}\sigma_{y}^{2}}p_{S}
\]
\[
\alpha_{0}=\frac{\alpha_{2}^{2}\sigma_{y}^{2}\mu_{x}-\alpha_{2}\alpha_{1}\sigma_{x}^{2}\mu_{y}}{\alpha_{1}^{2}\sigma_{x}^{2}+\alpha_{2}^{2}\sigma_{y}^{2}}=\mu\frac{\alpha_{2}^{2}\sigma_{y}^{2}-\alpha_{2}\alpha_{1}\sigma_{x}^{2}}{\alpha_{1}^{2}\sigma_{x}^{2}+\alpha_{2}^{2}\sigma_{y}^{2}}
\]
\[
\alpha_{1}=\frac{\alpha_{1}^{2}\sigma_{x}^{2}}{\alpha_{1}^{2}\sigma_{x}^{2}+\alpha_{2}^{2}\sigma_{y}^{2}}
\]
\[
\alpha_{2}=\frac{\alpha_{2}\alpha_{1}\sigma_{x}^{2}}{\alpha_{1}^{2}\sigma_{x}^{2}+\alpha_{2}^{2}\sigma_{y}^{2}}
\]
notice that if $\beta=\frac{1}{2}$ then at $t=0$ we have a solution
iff $\sigma$ are the same at $t=0$, 
\begin{align*}
\alpha_{0} & =0\\
\alpha_{1} & =\frac{1}{2}\\
\alpha_{2} & =\frac{1}{2}
\end{align*}
\[
\alpha_{0}=\frac{\alpha_{2}^{2}\sigma_{y}^{2}\mu_{x}-\alpha_{2}\alpha_{1}\sigma_{x}^{2}}{\alpha_{1}^{2}\sigma_{x}^{2}+\alpha_{2}^{2}\sigma_{y}^{2}}
\]
\[
\alpha_{1}=\frac{\alpha_{1}^{2}\sigma_{x}^{2}}{\alpha_{1}^{2}\sigma_{x}^{2}+\alpha_{2}^{2}\sigma_{y}^{2}}
\]
\[
\alpha_{2}=\frac{\alpha_{2}\alpha_{1}\sigma_{x}^{2}}{\alpha_{1}^{2}\sigma_{x}^{2}+\alpha_{2}^{2}\sigma_{y}^{2}}
\]
and 
\[
\mathbb{E}\left[\hat{p}|p\right]=\frac{\mathbb{E}\left[\frac{\alpha_{1}p+\alpha_{2}p_{S}-\alpha_{2}\mu_{y}}{\alpha_{1}}\right]\alpha_{1}^{2}\sigma_{x}^{2}+\alpha_{2}^{2}\sigma_{y}^{2}\mu_{x}}{\alpha_{1}^{2}\sigma_{x}^{2}+\alpha_{2}^{2}\sigma_{y}^{2}}=\frac{\alpha_{1}^{2}\sigma_{x}^{2}p+\alpha_{2}^{2}\sigma_{y}^{2}\mu_{x}}{\alpha_{1}^{2}\sigma_{x}^{2}+\alpha_{2}^{2}\sigma_{y}^{2}}
\]
and 
\[
\mathbb{E}\left[\hat{p}\right]=\frac{\alpha_{1}^{2}\sigma_{x}^{2}\mu_{x}+\alpha_{2}^{2}\sigma_{y}^{2}\mu_{x}}{\alpha_{1}^{2}\sigma_{x}^{2}+\alpha_{2}^{2}\sigma_{y}^{2}}=\mu_{x}
\]

\subsubsection*{Welfare Expressions}

The general formula is in the mathematica file, under the restriction $\mu_{x}=\mu_{y}$
and $\sigma_{x}=\sigma_{y}$ we get 
\[
W_{E}=-\left[\left(\alpha_{0}-\left(1-\alpha_{1}-\alpha_{2}\right)\mu\right)^{2}+\left(\left(1-\alpha_{1}\right)^{2}+\alpha_{2}^{2}\right)\sigma^{2}\right]
\]

We have welfare at $t=0$, where $\mu_{S}=1$. Namely 

\[
W_{E}=-\beta^{2}\left(\underbrace{\left(\mu_{x}-\mu_{y}\right)^{2}}_{\text{Prior Bias}}+\sigma_{x}^{2}+\sigma_{y}^{2}\right)
\]
\[
W_{P}=-\beta^{2}\frac{\sigma_{x}^{2}\sigma_{y}^{2}}{\sigma_{x}^{2}\left(1-\beta\right)^{2}+\sigma_{y}^{2}\beta^{2}}
\]
Then, 
\begin{align*}
\frac{W_{E}}{W_{P}} & =\frac{\left(\mu_{x}-\mu_{y}\right)^{2}+\sigma_{x}^{2}+\sigma_{y}^{2}}{\frac{\sigma_{x}^{2}\sigma_{y}^{2}}{\sigma_{x}^{2}\left(1-\beta\right)^{2}+\sigma_{y}^{2}\beta^{2}}}\\
\text{Assume equal \ensuremath{\sigma}} & =\frac{\left(\mu_{x}-\mu_{y}\right)^{2}+2\sigma^{2}}{\frac{\sigma^{2}}{\left(1-2\beta+2\beta^{2}\right)}}=\underbrace{\frac{\left(\mu_{x}-\mu_{y}\right)^{2}}{\frac{\sigma^{2}}{\left(1-2\beta+2\beta^{2}\right)}}}_{>0}+2\left(1-2\beta+2\beta^{2}\right)^{2}>2\left(\frac{1}{2}\right)=1
\end{align*}
so if $\mu_{x}=\mu_{y}$ (no prior bias), then $W_{E}\left(0\right)=W_{P}\left(0\right)$
iff $\sigma_{x}=\sigma_{y}$. 

\subsubsection*{Results}
$W_{E}>W_{P}$ this is because the bias/variance decomposition
\begin{align*}
W & =-\mathbb{E}\left[\left(p-\bar{p}\right)^{2}\right]=-\mathbb{E}\left[\left(\left(1-\mu_{T}-\mu_{A}\left[2\xi_{A}-1\right]-\mu_{S}\left(1-\beta\right)\right)p+\beta\mu_{S}p_{S}+\mu_{A}\left(1-\xi_{A}\right)\right)^{2}\right]\\
 & =-\mathbb{E}\left[\left(p-\hat{p}+\hat{p}-\bar{p}\right)^{2}\right]=-\left(\mathbb{E}\left[\left(p-\hat{p}\right)^{2}\right]+\mathbb{E}\left[\left(\hat{p}-\bar{p}\right)^{2}\right]+\xcancel{2\mathbb{E}\left[\left(p-\hat{p}\right)\left(\hat{p}-\bar{p}\right)\right]}\right)\\
 & =-\left(\underbrace{\mathbb{E}\left[\left(p-\hat{p}\right)^{2}\right]}_{\text{Precision of election}}+\underbrace{\mathbb{E}\left[\left(\hat{p}-\bar{p}\right)^{2}\right]}_{\text{Bias of elections}}\right)
\end{align*}
finally holds, the election have a bias.

The full characterization of the derivative (assuming equal $\mu$ and $\sigma$)
\[
\frac{\text{d}}{\text{d}t}W_{E}|_{t=0}=-4\beta\lambda_{1}\sigma^{2}\left(1-\beta-\xi_{A}\right)
\]
Instead assuming only equal $\mu$ we have 
\[
\frac{\text{d}}{\text{d}t}W_{E}|_{t=0}=-2\beta\lambda_{1}\left(\beta\sigma_{y}^{2}-\sigma_{x}^{2}\left(2\left(1-\xi_{A}\right)-\beta\right)\right)
\]
so 
\[
\frac{\text{d}}{\text{d}t}W_{E}|_{t=0}>0\iff1-\beta<\xi_{A}
\]
or in general 
\[
\frac{\beta}{2\left(1-\xi_{A}\right)-\beta}<\frac{\sigma_{x}^{2}}{\sigma_{y}^{2}}
\]
a sensible condition. Also, $\lambda_{1}$ magnifies either the positive
or the negative change local to $0$ and in particular if $1-\beta>\xi_{A}$
then more $\lambda_{1}$ is bad for welfare local to $t=0$. To the
contrary, 
\[
\frac{\text{d}}{\text{d}t}W_{P}|_{t=0}=\frac{2\left(1-\beta\right)\beta^{2}\lambda_{1}\sigma^{2}\left(2\xi_{A}-1\right)}{\text{sthg}^{2}}>0
\]
 and the welfare of the unconstrained principal is {[}but this is
just a conjecture not falsified by Math plots{]} always increasing
in both $\lambda_{1},t$.

\newpage
\textbf{Conjecture}

$W_{P}$ is increasing in $t$ (and $\lambda_{1}$)--- We show that
\begin{align*}
\frac{\text{d}}{\text{d}t}W_{P} & \propto-\left[\underbrace{2\left(1-\xi_{A}\right)\mu_{S}}_{+}\underbrace{\frac{\text{d}}{\text{d}t}\mu_{A}}_{?}+\underbrace{\left(1-2\left(1-\xi_{A}\right)\mu_{A}\right)}_{+}\underbrace{\frac{\text{d}}{\text{d}t}\mu_{S}}_{-}\right]\\
& =\underbrace{\exp\left\{ -\left(\lambda_{1}+\lambda_{2}\right)t\right\} }_{+}\lambda_{1}\left(2\left(1-\xi_{A}\right)-\exp\left\{ \lambda_{2}t\right\} \right)
2\left(1-\xi_{A}\right)-\exp\left\{ \lambda_{2}t\right\} <2\left(1-\xi_{A}\right)-1\\
& =1-2\xi_{A}<0
\end{align*}

when computed in 
\[
\frac{\text{d}}{\text{d}\lambda_{1}}W_{P}\propto-\left[\underbrace{2\left(1-\xi_{A}\right)\mu_{S}}_{+}\underbrace{\frac{\text{d}}{\text{d}\lambda_{1}}\mu_{A}}_{?}+\underbrace{\left(1-2\left(1-\xi_{A}\right)\mu_{A}\right)}_{+}\underbrace{\frac{\text{d}}{\text{d}\lambda_{1}}\mu_{S}}_{-}\right]
\]
which has the same sign as

\[
\frac{\text{d}}{\text{d}\lambda_{1}}W_{P} \propto-\exp\left\{ \lambda_{2}t\right\} \lambda_{2}^{2}t+2\lambda_{2}\left[\left(\exp\left\{ \lambda_{2}t\right\} \lambda_{1}t\right)+\left(1-\xi_{A}\right)\exp\left\{ \left(\lambda_{2} -\lambda_{1}\right)t\right\} -\left(1-\xi_{A}\right)\left(1+\lambda_{1}t\right)\right] -\lambda_{1}^{2}t\left(\exp\left\{ \lambda_{2}t\right\} -2\left(1-\xi_{A}\right)\right)
\]

\bigskip

Furthermore, analyse $\xi_{A}$:
\begin{align*}
\xi_{A} \\ 
& =-\exp\left\{ \lambda_{2}t\right\}
\lambda_{2}^{2}t+2\lambda_{2}\left[\left(\exp\left\{ \lambda_{2}t\right\} \lambda_{1}t\right)\right]-\lambda_{1}^{2}t\left(\exp\left\{ \lambda_{2}t\right\} \right)\\
& =-t\exp\left\{ \lambda_{2}t\right\} \left(\lambda_{2}-\lambda_{1}\right)^{2}+2\lambda_{2}\left(1-\xi_{A}\right)\left[\exp\left\{ \left(\lambda_{2}-\lambda_{1}\right)t\right\} -\left(1+\lambda_{1}t\right)\right]+2\lambda_{1}^{2}t\left(1-\xi_{A}\right)\\
& =\underbrace{-t\exp\left\{ \lambda_{2}t\right\} \left(\lambda_{2}-\lambda_{1}\right)^{2}}_{\textit{negative}}+2\lambda_{2}\left(1-\xi_{A}\right)\left[\exp\left\{ \left(\lambda_{2}-\lambda_{1}\right)t\right\} -\left(1+\lambda_{1}t\right)+2\lambda_{1}^{2}t\right]
\end{align*}
 
Now if the second addendum is negative then we are done; so assume
it is positive, that is 
\begin{align*}
\exp\left\{ \left(\lambda_{2}-\lambda_{1}\right)t\right\} -\left(1+\lambda_{1}t\right)+2\lambda_{1}^{2}t>0
\end{align*}

 then the sum is smaller than 

\begin{align*}
\underbrace{-t\exp\left\{ \lambda_{2}t\right\} \left(\lambda_{2}-\lambda_{1}\right)^{2}}_{<0}+\lambda_{2}\left[\exp\left\{ \left(\lambda_{2}-\lambda_{1}\right)t\right\} \left(1+\lambda_{1}t\right)+2\lambda_{1}^{2}t\right] \\ 
& =
\lambda_{1}^{2}t-\exp\left\{ \lambda_{2}t\right\} \left(\lambda_{2}-\lambda_{1}\right)^{2}t+\lambda_{2}\left(-1+\exp\left\{ \left(\lambda_{2}-\lambda_{1}\right)t\right\} -\lambda_{1}t\right) \\ 
& = \lambda_{1}^{2}t+\lambda_{2}\exp\left\{ \left(\lambda_{2}-\lambda_{1}\right)t\right\} -\left[\exp\left\{ \lambda_{2}t\right\} \left(\lambda_{2}-\lambda_{1}\right)^{2}t+\lambda_{2}\left(1+\lambda_{1}t\right)\right] <0
\end{align*}

so it remains to show that this is always negative; if $\lambda_{1}\approx0$

\begin{align*}
-\lambda_{2}\left(1-\exp\left\{ \lambda_{2}t\right\} \left(1-\lambda_{2}t\right)\right)<-\lambda_{2}^{2}t<0
\end{align*}

$W_{E}$ has interesting comparative statistics due to the interaction with bias.
In particular, it seems that for $\beta>\text{stgh}$, then {[}if
there is no prior bias, $\mu_{x}=\mu_{y}${]} there is a time $t$
such that $Bias\left(t\right)=0$ because the evolution of $\mu_{S},\mu_{A}$
is such that $\alpha^{E}=\alpha^{P}$. This seems interesting, possibly
a result to put in a proposition.


Based on our model, we can draw three main results and one additional
interesting result.

\paragraph{Inefficiency of twitter economy and non-monotonicity in election
times.}

The first result relates to the political institutions of the digital
economy. From our model is that a Twitter-Facebook economy where everyone
can speak their mind is not necessarily good: indeed we want only
those that went through critical thinking to vote. Following this point,
the naturally arising question, \emph{when do we want to hold elections}?
Our model clearly implies non-monotonicity in time for election periods.

\paragraph{Typology of voting-users and adverse selection.}

The second result relates to the typology of voting-users. The ``\emph{clients''
of news outlets}, in a micro-foundation of the $\lambda$ functions
are either low $i$ partisans (that look at it for fun) or frustrated
critical thinking voting-users that look for some facts (positive predictions).
On a related but different point, we can identify the \emph{adverse
selection in the vote-force} (under some conditions the strengths
of the stereotype pool weakens), and how the  format amplifies
/ reduces this issue (always true that it is better if only types
vote, at least in the symmetric case).

\paragraph{Partisan format and compensation effect.}

The third and most intriguing result relates to the  format.
We can study the \emph{impact of different storytelling formats} (more
in depth, helps the high $i$, but how it correlates with $\alpha$):
more in depth, but keep it somehow primitive. In particular and more
interestingly, we can allow for \emph{asymmetries}: either there is
a ``better'' policy (say $\beta=1$, so upon reflecting everyone
agrees $1$ is right), or stereotypes of one side are less likely
to enter critical thinking (evidence that conservatives are overconfident),
how does this change the outcome, as well as the incentives for the
critical thinking agents (that may vote for those that are less confident
because of the bias in the type pool). The problem of asymmetries
is that a partisan  format, or is the fact that one stereotype
is more attractive than the other to  make the problem of agents in critical thinking more problematic: remember they are smart
but unwise, so they cannot ignore the fact of a stereotyped partisan pool, either because stereotypes are more resistant, or because the
shifts the stereotypes. Hence, we propose to explain such a situation by an effect that we label the "Compensation Effect": When you perceive the  device to be partisan in one direction, you vote in the opposite direction when in critical thinking.

\paragraph{The benefits of making voting costly.}
A resulting and potentially controversial consequence of such an asymmetry
is that \textit{voting costs} in this situation may be positive because
they can also exclude the strategic types that recognize the stereotype pool is partisan and cannot morally abstain or vote against their type. They can use the excuse not to go to vote.

\subsection{Proofs}

\section{Experiment With A Three Cognitive-State Model}

In the experiment, we gathered data to decompose the critical thinking process into three stages: ${S, A, T}$. Here, $S$ remains unchanged. Now, $A$ denotes an intermediate, transitory stage during which agents experience internal uncertainty regarding the formation of their stable preferences. Finally, $T$  
denotes the stage in which agents have completed the critical thinking process and have formed their stable preferences.

\bigskip
\bigskip

\begin{adjustbox}{max totalsize={1\textwidth}{1\textheight},center}
\footnotesize
\begin{tikzpicture}[font=\footnotesize,every node/.style={align=center}]
		\draw[] (0,0)--(23.3,0);
		\foreach \i[count=\j] in {.2,1.5,...,10,14,15.3,...,24}
		\draw (\i,.15)--coordinate[pos=.5] (c\j) (\i,-.15);
		\node[above=3mm, align=left] (nc) at ($(c8)!.55!(c9)$) {
			$\bullet\,\,$ Short \& Crude\\
			$\bullet\,\,$ Medium \& Reasoned\\
			$\bullet\,\,$ Long \& Reasoned\\
			$\bullet\,\,$ Short, Bias \& Crude
		};
		\node[rotate=90,yshift=3mm] at (nc.west) {randomized};
		\draw [green!70!black,decorate,decoration = {brace,mirror, raise=5pt, amplitude=5pt}] ($(c1)+(0mm,-1.9cm)$) -- node[below=5mm,align=center] {$Part 1$} ($(c3)+(-0.1cm,-1.9cm)$);
		\draw [green!70!black,decorate,decoration = {brace,mirror, raise=5pt, amplitude=5pt}] ($(c3)+(0.1cm,-1.9cm)$) -- node[below=5mm,align=center] {$Part 2$} ($(c7)+(-0.1cm,-1.9cm)$);
		\draw [green!70!black,decorate,decoration = {brace,mirror, raise=5pt, amplitude=5pt}] ($(c8)+(0.1cm,-1.9cm)$) -- node[below=5mm,align=center] {$Part3$} ($(c9)+(-0.1cm,-1.9cm)$);
		
		\draw [green!70!black,decorate,decoration = {brace,mirror, raise=5pt, amplitude=5pt}] ($(c9)+(0.1cm,-1.9cm)$) -- node[below=5mm,align=center] {$Part 4$} ($(c11)+(-0.1cm,-1.9cm)$);
		
		\draw [green!70!black,decorate,decoration = {brace,mirror, raise=5pt, amplitude=5pt}] ($(c12)+(0.1cm,-1.9cm)$) -- node[below=5mm,align=center] {$Part 5$} ($(c15)+(-0.1cm,-1.9cm)$);
		
		\draw [green!70!black,decorate,decoration = {brace,mirror, raise=5pt, amplitude=5pt}] ($(c15)+(0.1cm,-1.9cm)$) -- node[below=5mm,align=center] {$Part6$} ($(c16)+(-0.1cm,-1.9cm)$);
  
		\draw [decorate,decoration = {brace, raise=5pt, amplitude=5pt}] ($(c3)+(0.1cm,2.5cm)$) -- node[above=5mm,align=center] (nn1) {4-step test} ($(c7)+(-0.1cm,2.5cm)$);
		
		\draw [decorate,decoration = {brace, raise=5pt, amplitude=5pt}] ($(c8)+(0.1cm,2.5cm)$) -- node[above=5mm,align=center] (nn2) {4 storytelling formats} ($(c9)+(-0.1cm,2.5cm)$);
		
		\draw [decorate,decoration = {brace, raise=5pt, amplitude=5pt}]  ($(c9)+(0.1cm,2.5cm)$) -- node[above=5mm,align=center] (nn3) {Cooling Period} ($(c11)+(-0.1cm,2.5cm)$);
		
		\draw [decorate,decoration = {brace, raise=5pt, amplitude=5pt}] ($(c12)+(0.1cm,2.5cm)$) -- node[above=5mm,align=center] (nn4) {3-step test} ($(c15)+(-0.1cm,2.5cm)$);
  
		\draw[stealth'-] ($(nn1)+(0,.3)$)-- node[pos=1,above] {Elicitation of\\Ex-Ante Types} ++(0,.7);
		\draw[stealth'-] ($(nn2)+(0,.3)$)-- node[pos=1,above] {Treatment} ++(0,.7);
		\draw[stealth'-] ($(nn3)+(0,.3)$)-- node[pos=1,above] {Elicitation of \\Cognitive Styles} ++(0,.7);
		\draw[stealth'-] ($(nn4)+(0,.3)$)-- node[pos=1,above] {Elicitation of \\Ex-Post Types} ++(0,.7);
		\coordinate (f1) at ($(c3)!.6!(c4)$);
		\draw[stealth'-, blue!70!teal] ($(f1)+(-.05,-.3)$)-- node[pos=1,below] {{\bf \textcolor{blue}{Incentivized}}} ++(.0,-.5);
		\coordinate (f2) at ($(c12)!.6!(c13)$);
		\draw[stealth'-, blue!70!teal] ($(f2)+(-.05,-.3)$)-- node[pos=1,below] {{\bf \textcolor{blue}{Incentivized}}} ++(0,-.5);
		\def\angle{85}
		\def\size{\footnotesize}
		\node[rotate=\angle, anchor=west, xshift=2mm, text width=2cm, font=\size] at ($(c1)!.5!(c2)$) {Demographics};
		\node[rotate=\angle, anchor=west, xshift=2mm, text width=2cm, font=\size] at ($(c2)!.5!(c3)$) {Ex-Ante Preferences};
		\node[rotate=\angle, anchor=west, xshift=2mm, text width=2cm, font=\size] at ($(c3)!.5!(c4)$) {\bf \textcolor{blue}{Knowledge Test}};
		\node[rotate=\angle, anchor=west, xshift=2mm, text width=2cm, font=\size] at ($(c4)!.5!(c5)$) {Listing Reasons};
		\node[rotate=\angle, anchor=west, xshift=2mm, text width=2cm, font=\size] at ($(c5)!.5!(c6)$) {Issue Familiarity};
		\node[rotate=\angle, anchor=west, xshift=2mm, text width=2cm, font=\size] at ($(c6)!.5!(c7)$) {\textcolor{red}{Internal Uncertainty}};
		\node[rotate=\angle, anchor=west, xshift=2mm, text width=2cm, font=\size] at ($(c9)!.5!(c10)$) {Need For Cognition};
		\node[rotate=\angle, anchor=west, xshift=2mm, text width=2cm, font=\size] at ($(c10)!.5!(c11)$) {Cognitive Flexibility};
		\node[rotate=\angle, anchor=west, xshift=2mm, text width=2cm, font=\size] at ($(c12)!.5!(c13)$) {\bf \textcolor{blue}{Essay Writing}};
		\node[rotate=\angle, anchor=west, xshift=2mm, text width=2cm, font=\size] at ($(c13)!.5!(c14)$) {Ex-Post Preferences};
		\node[rotate=\angle, anchor=west, xshift=2mm, text width=2cm, font=\size] at ($(c14)!.5!(c15)$) {\textcolor{red}{Internal Uncertainty}};
		\node[rotate=\angle, anchor=west, xshift=2mm, text width=2cm, font=\size] at ($(c15)!.5!(c16)$) {News Habits};
		
		\draw[dotted, red] ([yshift=5.35cm]c3) rectangle ([yshift=4.2cm]c7);
  	\draw[dotted, red] ([yshift=5.35cm]c9) rectangle ([yshift=4.2cm]c11);
		\draw[dotted, red] ([yshift=5.35cm]c12) rectangle ([yshift=4.2cm]c15);
\end{tikzpicture}
\end{adjustbox}

\bigskip
\bigskip

Table \ref{tab:classstrat} shows the classification strategy of participants as \textit{S}tereotype, \textit{A}ware, and \textit{T}ype.

 \begin{table}[H]
        \centering
        \footnotesize
        \addtolength{\tabcolsep}{5pt} 
        \begin{tabular}{llll}
        \toprule \toprule
        Treatment &\multicolumn{1}{c}{$ T $} & \multicolumn{1}{c}{$ A $} & \multicolumn{1}{c}{$ S $} \\
        \midrule
        &  Knowledge Test Score $>$ $\tau_{KTS}$ \\
\sc{before}  & Issue Familiarity = 1  & Knowledge Test Score $>$ $\tau_{KTS}$ & \\
        & \textcolor{red}{\bf Internal Uncertainty $\neq 0$}\\
                & Reasons List $>$ $\tau_{RL}$   \\
                & & Issue Familiarity = 1 \\ 
        \midrule
        \sc{after}    & Psychologists Grade = Pass & Else  &   \\
        \bottomrule \bottomrule
        \end{tabular}
        \caption{\sc Classification \textbf{strategy} before/after treatment}
        \label{tab:classstrat}
\end{table}

\bigskip
\bigskip

The analysis presents the frequencies of the three states of participants before and after the treatment

 \begin{table}[H]
        \centering
        \footnotesize
        \addtolength{\tabcolsep}{5pt} 
        \begin{tabular}{llll}
        \toprule \toprule
        & \multicolumn{1}{c}{$ S_{1} $} & \multicolumn{1}{c}{$ A_{1} $} & \multicolumn{1}{c}{$ T_{1} $} \\
        \midrule
            $S_{0}$   &  475 & 153 & 39 \\
            $A_{0}$ & 0 & 21 & 2 \\
            $T_{0}$ & 0 & 0 & 30 \\
        \bottomrule \bottomrule
        \end{tabular}
        \caption{\sc Table: Frequencies before/after treatment}
        \label{tab:classstrat3}
\end{table}
\end{document}